\documentclass[12pt]{article}

\usepackage{graphicx}
\usepackage{fancybox}
\usepackage{amsmath}
\usepackage{amssymb}
\usepackage{latexsym}
\usepackage{epsfig}

\newcommand{\om}{\omega}
\newcommand{\al}{\alpha}
\newcommand{\ep}{\epsilon}

\newcommand{\la}{\lambda}
\newcommand{\La}{\Lambda}

\newcommand{\NS}{\mbox{NS}}
\newcommand{\tNS}{\widetilde{\mbox{NS}}}
\newcommand{\R}{\mbox{R}}
\newcommand{\tR}{\widetilde{\mbox{R}}}

\newcommand{\stR}{\widetilde{\msc{R}}}

\newcommand{\del}{\partial}

\newcommand{\msc}[1]{\mbox{\scriptsize #1}}
\newcommand{\dsp}{\displaystyle}

\newcommand{\bc}{\Bbb C}
\newcommand{\br}{\Bbb R}
\newcommand{\bz}{\Bbb Z}

\newcommand{\bh}{{\Bbb H}}

\newcommand{\bsz}{\Bbb Z}

\newcommand{\bm}[1]{\mbox{\boldmath ${#1}$}}

\newcommand{\cJ}{{\cal J}}

\newcommand{\cN}{{\cal N}}

\newcommand{\cC}{{\cal C}}

\newcommand{\cZ}{{\cal Z}}

\newcommand{\tcZ}{\widetilde{\cZ}}

\newcommand{\hc}{\hat{c}}

\newcommand{\hf}{\widehat{f}}

\newcommand{\Z}{\bm{Z}}

\newcommand{\Th}[2]{\Theta_{#1,#2}}
\renewcommand{\th}{{\theta}}

\newcommand{\ch}[2]{\mbox{ch}^{#1}_{#2}}

\newcommand{\hch}[2]{\widehat{\mbox{ch}}^{#1}_{#2}}

\newcommand{\chd}{\mbox{ch}_{\msc{dis}}}
\newcommand{\chg}{\mbox{ch}_{\msc{grav}}}
\newcommand{\hchd}{\widehat{\mbox{ch}}_{\msc{dis}}}
\newcommand{\hchg}{\widehat{\mbox{ch}}_{\msc{grav}}}

\newcommand{\chig}{\chi_{\msc{grav}}}
\newcommand{\chid}{\chi_{\msc{dis}}}

\newcommand{\hchid}{\widehat{\chi}_{\msc{dis}}}

\newcommand{\hchig}{\widehat{\chi}_{\msc{grav}}}

\newcommand{\tpsi}{\tilde{\psi}}

\newcommand{\erf}{\mbox{Erf}}

\newcommand{\sgn}{\mbox{sgn}}

\newcommand{\nn}{\nonumber\\}

\renewcommand{\Im}{{\rm Im}}

\newcommand{\vth}{\vartheta}

\def\boxit#1{\vbox{\hrule\hbox{\vrule\kern8pt
\vbox{\hbox{\kern8pt}\hbox{\vbox{#1}}\hbox{\kern8pt}}
\kern8pt\vrule}\hrule}}
\def\mathboxit#1{\vbox{\hrule\hbox{\vrule\kern8pt\vbox{\kern8pt
\hbox{$\displaystyle #1$}\kern8pt}\kern8pt\vrule}\hrule}}

\newcommand{\any}{{}^{\forall}}

\renewcommand{\mod}{\, \mbox{mod} ~ }

\newcommand {\eqn}[1]{(\ref{#1})}

\makeatletter
\@addtoreset{equation}{section}
\def\theequation{\thesection.\arabic{equation}}
\makeatother

\begin{document}

\begin{titlepage}
 \
 \renewcommand{\thefootnote}{\fnsymbol{footnote}}
 \font\csc=cmcsc10 scaled\magstep1
 {\baselineskip=14pt
 \rightline{
 \vbox{\hbox{\today}
       }}}

 \baselineskip=20pt
 
\begin{center}

{\bf \Large  
Compact Formulas for the \\
Completed Mock Modular Forms

}

 
\vskip 2cm

\noindent{ \large Tohru Eguchi}\footnote{\sf tohru.eguchi@gmail.com}
 \\

\medskip

{\it Department of Physics and Research Center for Mathematical Physics, \\
Rikkyo University, Tokyo 171-8501, Japan}

\vskip 8mm
\noindent{ \large Yuji Sugawara}\footnote{\sf ysugawa@se.ritsumei.ac.jp}
\\

\medskip

 {\it Department of Physical Sciences, 
 College of Science and Engineering, \\ 
Ritsumeikan University,  
Shiga 525-8577, Japan}
 

\end{center}

\bigskip

\begin{abstract}

In this paper we present a new compact expression of the elliptic genus of 
$SL(2)/U(1)$-supercoset theory by making use of  the `spectral flow method' 
of the path-integral evaluation.
This new expression is written in a form like a Poincar\'e series
with a non-holomorphic Gaussian damping factor, and manifestly shows
the modular and spectral flow properties of a real analytic Jacobi form. 
As a related problem, we present similar compact formulas for  the modular completions of various mock modular forms which appear in the representation theory of
  $\cN=2,4$ superconformal algebras.
  
We further discuss the generalization to the cases of arbitrary spin-structures, that is, 
the world-sheet fermions with twisted boundary conditions parameterized by a continuous parameter.
This parameter is naturally identified with the `$u$-variable' in the Appell-Lerch sum.

\end{abstract}


\setcounter{footnote}{0}
\renewcommand{\thefootnote}{\arabic{footnote}}

\end{titlepage}

\baselineskip 18pt

\vskip2cm 


\section{Introduction and Summary}

~

It is known that superconformal field theories with non-compact target space possess interesting theoretical issues; for instance, in order to regulate its IR divergences  we may sometimes have to sacrifice its holomorphy property.
A good laboratory to study them is the $SL(2)/U(1)$-supercoset theory, which is interpreted as the supersymmetric extension of 
the model of 2-dimensional black hole \cite{2DBH}. 
One of the important progress in the study of this model 
has been the appearance of non-holomorphicity in the elliptic genus  
\cite{Troost,ES-NH,AT}. This originates from the existence of gapless continuous spectrum of 
non-BPS states causing the IR-divergence in the theory. 
Quite interestingly, these studies 
have revealed the physical origin of a mathematical construction of  `modular completions' of mock modular forms \cite{Zwegers}. 
Namely, it turned out that elliptic genera of $SL(2)/U(1)$-theory \cite{Troost, ES-NH}
are written in terms of the modular completion 
of the mock modular form  \cite{Zwegers};
\begin{equation}
\widehat{f}_u^{(k)}(\tau,z)
:=   
f^{(k)}_u(\tau,z)
- \frac{1}{2} \sum_{m\in \bz_{2k}} \, R_{m,k}(\tau, u) \Th{m}{k} (\tau, 2z), 
\hspace{1cm} (k\in \bz_{>0}).
\label{hAppell}
\end{equation}
Here the
mock modular form (Appell-Lerch sum) is defined as 
\begin{equation}
f^{(k)}_u(\tau,z)
:=  \sum_{n\in \bsz} 
\frac{q^{ k n^2} y^{2 kn} }{1-y w^{-1} q^n} ,
\hspace{1cm} 
(q\equiv e^{2\pi i\tau}, ~ y\equiv e^{2\pi i z}, ~ w\equiv e^{2\pi i u} ),
\label{Appell}
\end{equation}
and has an anomalous modular transformation law
(this function with $u=0$ is closely related to the character of massless representations of ${\mathcal N}=4$ superconformal algebra \cite{ET}).
Anomalous transformation law is compensated by the second term in the R.H.S. of (\ref{hAppell}) with  
\begin{eqnarray}
&&R_{m,k}(\tau,u) := \sum_{\nu \in m+ 2k \bz} \,
\left[ \sgn(\nu+0) -  
\erf\left\{ \sqrt{\frac{\pi \tau_2}{k}} \left(\nu+ 2k \frac{u_2}{\tau_2} 
\right)\right\} \right]\,
w^{-\nu} q^{-\frac{\nu^2}{4k}},
\nn
&&
\hspace{9cm}
\left(\tau_2\equiv \Im \, \tau, ~ u_2 \equiv \Im \, u\right)
\label{Rmk}
\end{eqnarray}
which has a non-holomorphic dependence on $\tau$. Completed mock modular form $\hat{f}_u^{(k)}(\tau,z)$ has a well-defined transformation law as a Jacobi form of weight $1$ and index $k$.
(See Appendix A for the convention of error function $\erf(x)$, theta function $\Th{m}{k}(\tau,z)$ and Jacobi forms.)
This means that the elliptic genus of $SL(2)/U(1)$-supercoset is described by  
a non-holomorphic generalization of  a Jacobi form.


Moreover, the torus partition function has also been analyzed in \cite{ES-NH}, and is shown 
to be expanded in terms of the modular completions of extended discrete characters (see Appendix D) as well as 
continuous (non-BPS) characters. 
More recently, in literatures \cite{AT2,Murthy}, the elliptic genera of the $SL(2)/U(1)$-theory 
as well as some generalized models have been analyzed 
using the formulation of the gauged linear sigma model as proposed in \cite{HK}.


~

In this paper,  we shall propose  a method of deriving compact expressions 
for the completed mock modular forms {\em that makes their modular properties manifest.} 
To this end, we adopt the technique of {\em spectral flow method},
which was partly studied in \cite{orb-ncpart}, 
rather than the method of character decomposition in \cite{ES-NH}. 
Using the new method, we can simplify relevant calculations. 

What is important  is that we are naturally led to an expression of  elliptic genus or completed mock modular forms 
as a sum over a 2-dimensional lattice 
$$
\cZ(\tau,z) \sim \sum_{m,n\in \bz} \, g_{(m,n)} (\tau,z) \left( \equiv \sum_{\la \equiv m\tau+n \in \La}\, g_{\la} (\tau,z), ~~~ (\La \equiv \bz \tau + \bz)
\right),
$$
where $g_{(m,n)} (\tau,z)$ behaves covariantly under modular  transformations;
$$
g_{(m,n)}(\tau+1, z) = g_{(m,m+n)} (\tau,z) , \hspace{1cm} 
g_{(m,n)} \left( -\frac{1}{\tau}, \frac{z}{\tau} \right) = e^{i\pi \frac{\hc}{\tau} z^2} \, 
g_{(n,-m)} (\tau,z).
$$
($\hat{c}$ is defined in (\ref{chat})).
Therefore, this expression manifestly exhibits modular invariance   
as in the case of Poincar\'e or Eisenstein series. 
The two integers $m$, $n$ are identified with spectral parameters 
along the spatial and temporal cycles of a torus.

~


Let us summarize the main results of this paper; 
\begin{description}
\item[(i)] 
The main formula  we will derive is given in
\eqn{EG nh-E 2}. This is  written  
in the form of non-holomorphic Poincar\'e type series mentioned above,
and will be a most tranparent  expression of elliptic genus of the cigar model.
We will reach this formula in two ways: One is a direct evaluation of path-integration 
with the help of spectral flow expansion. The other is based on the properties of 
the modular completions of extended or irreducible discrete characters  
given in the previous papers \cite{ES-NH,orb-ncpart}.
Combining these two approaches, we can rewrite  
the function $\hf^{(k)}_{u=0}(\tau,z)$ \eqn{hAppell} as given in \eqn{hAppell nh-E}.
We can also express the modular completions of discrete characters  
in the forms of double series mentioned above. 
These are given in \eqn{hchid nh-E} and \eqn{hchd nh-E}.

\item[(ii)]
We also discuss the generalization to the cases of arbitrary spin-structures, that is, 
the world-sheet fermions with twisted boundary conditions parameterized by a continuous parameter.
This continuous parameter is naturally identified with the `$u$-variable' in the Appell-Lerch sum.    
The complete formula of rewriting  $\hf^{(k)}_{u}(\tau,z)$ in terms of the non-holomorphic Poincar\'e series 
is given in \eqn{hAppell nh-E u}.   

\end{description}

~

~


\section{Revisiting Elliptic Genus of $SL(2)/U(1)$-supercoset model  as 
a Non-holomorphic Poincar\'e Series}

~


\subsection{A Refined Calculation of Elliptic 
Genus Based on the Spectral Flow Method}


We shall start with the torus partition function with the general twist angles $z, \bar{z} \in \bc$
(in the $\tR$-sector) of the cigar $SL(2)/U(1)$-supercoset model, which has been evaluated in \cite{ES-NH}\footnote
   {See also \cite{HPT,IPT,ES-BH} for the earlier works in which closely related 
  analyses are presented. }. 
Throughout this paper we set the level of $SL(2)$ super-WZW model to be a real positive number $k$ (the level of bosonic part
is $k+2$), which means that the central charge of this superconformal system is
\begin{equation}
\hc \left(\equiv \frac{c}{3} \right) = 1 + \frac{2}{k}.
\label{chat}
\end{equation}
Because of the IR-divergence of cigar geometry, 
we have to introduce a suitable regularization. 
Here we shall adopt a `refined' scheme of regularization.
Namely, we start with the following regularized partition function\footnote
{We shall denote the regularized partition function as `$\Z_{\msc{reg}}(\tau,z, \bar{z}; \ep)$'
rather than `$\Z_{\msc{reg}}(\tau,z ; \ep)$' although $\bar{z}$ is just the complex conjugate of $z$ 
in \eqn{Z reg}. This is because we will later treat $z$ and $\bar{z}$ as two {\em independent\/} complex
variables to derive the elliptic genus, while $\bar{\tau}$ is always the complex conjugate of $\tau$ 
through this paper.
}
($\mu\equiv \mu_1+i\mu_2, ~ z\equiv z_1 +i z_2,~~ \mu_1, \mu_2,  z_1, z_2 \in \br$) 
\begin{eqnarray}
\Z_{\msc{reg}}(\tau,z, \bar{z}; \ep) &=&  
k e^{2\pi \frac{\hc}{\tau_2} |z|^2 - 2\pi \frac{k+4}{k\tau_2} z_2^2} \, 
\int_{\bc} \frac{d^2 \mu}{\tau_2}\, 
\sigma(\tau, \mu,z, \bar{z}; \ep)\,
\left|
\frac{\th_1\left(\tau, \mu + \frac{k+2}{k}z\right)}
{\th_1\left(\tau, \mu + \frac{2}{k} z \right)}\right|^2 
\, e^{- 4\pi  z_2 \frac{\mu_2}{\tau_2}}\,
e^{-\frac{\pi k}{\tau_2}\left| \mu \right|^2},
\nn
&&
\label{Z reg}
\end{eqnarray}
where we introduced a regularization factor 
$\sigma(\tau, \mu,z, \bar{z};\ep)$ 
($\ep >0$)
defined as
\begin{eqnarray}
\sigma(\tau, \mu,z,\bar{z};\ep) := \prod_{m_1,m_2 \in \bz} \, \left[ 1 - 
 e^{-\frac{1}{\ep \tau_2} 
\left\{ (s_1+m_1) \tau+ (s_2+m_2) + \frac{2z}{k} \right\}
\left\{ (s_1+m_1) \bar{\tau}+ (s_2+m_2) + \frac{2\bar{z}}{k} \right\}
}\right],
\label{def sigma}
\end{eqnarray}
with $\mu \equiv s_1 \tau+ s_2,\, s_i \in \br, i=1,2$.
Note that all the singularities of the integrand located at $\mu + \frac{2}{k} z \in \bz \tau + \bz$
that originate from the $\th_1$-factor 
are removed by inserting  \eqn{def sigma}
and the $\mu$-integral converges as long as $\ep>0$.
We will later regard  $\sigma(\tau, \mu, z, \bar{z} ;\ep)$ as a holomorphic function
with respect to {\em complex\/} variables $s_1, s_2$.

Let us very briefly sketch how the torus partition function \eqn{Z reg} is derived. 
See \cite{ES-NH} for the details; 
\begin{itemize}
\item 
According to the standard analysis of gauged WZW models \cite{GawK,Schnitzer,KarS}, 
the relevant calculation reduces to that of free fields, except for 
the integral of modulus $\mu$ that is coupled to the 
anomaly-free $U(1)$-current we should gauge;
\begin{equation}
\cJ \equiv j^3 + \psi^+ \psi^- + i\sqrt{k} \del Y ,
\label{gauge current}
\end{equation}
where $j^3$ denotes the contribution from the bosonic $SL(2)$-WZW with level $k+2$ and $\psi^+$, $\psi^-$ are the world-sheet fermions.
The compact boson $Y$ corresponds to the degrees of freedom of gauge field which cannot be gauged away.
This obeys  the twisted boundary condition parameterized by the $\mu$-variable, and 
the Gaussian factor $e^{-\frac{\pi k}{\tau_2} \left|\mu\right|^2} $ 
is the contribution of its zero-modes. 
Note that the range of $\mu $-integral is extended from 
$\mbox{Jac} (\Sigma) \cong \Sigma \cong \bc/ \La$, ($\La \equiv \bz \tau + \bz$) 
to the whole complex plane  due to the winding modes of $Y$.

\item
The $z$-parameter (`chemical potential') couples to 
the left-moving $\cN=2$ $U(1)$-current given by \cite{KS}
\begin{equation}
J = 
\psi^+ \psi^- + \frac{2}{k} 
\left( j^3 + \psi^+  \psi^- \right) 
\equiv \frac{k+2}{k} \psi^+  \psi^- + \frac{2}{k} j^3.
\label{N=2 U(1)}
\end{equation}
Notice that the Gaussian factor 
$e^{-\frac{\pi k}{\tau_2} \left|\mu\right|^2} $ includes no dependence 
on $z$ in the torus partition function \eqn{Z reg}. 
This should be the case, 
since  $J$ possesses no contribution from the boson $Y$. 

\end{itemize}

~

Now, the $\mu$-integral in  
\eqn{Z reg} is finite  as long as $\ep >0$,
and because of the modular invariance 
of $\sigma(\tau, \mu,z,\bar{z} ;\ep)$, 
$\Z_{\msc{reg}}(\tau,z, \bar{z}; \ep)$
is defined 
so as to preserve the modular property manifestly;
\begin{equation}
\Z_{\msc{reg}}(\tau+1,z, \bar{z}; \ep) = \Z_{\msc{reg}}(\tau,z, \bar{z}; \ep) ,
\hspace{1cm} 
\Z_{\msc{reg}}\left(-\frac{1}{\tau},\frac{z}{\tau}, \frac{\bar{z}}{\bar{\tau}}; \ep\right) 
= \Z_{\msc{reg}}(\tau,z, \bar{z}; \ep). 
\end{equation}

The partition function $\Z_{\msc{reg}}(\tau,z, \bar{z}; \ep) $
logarithmically diverges in  the limit $\ep\, \rightarrow\, +0$, 
and the divergent part is identified with the contributions from  
the strings propagating in the asymptotic cylindrical region of the cigar.
This implies that the characteristic behavior around $\ep \, \rightarrow \, +0$ is given by 
\begin{eqnarray}
&& \Z_{\msc{reg}}(\tau, z, \bar{z}; \ep) = \cC 
| \ln \ep |
e^{2\pi \frac{\hc}{\tau_2} z_1^2}\,
\left|\frac{\th_1(\tau,z)}{\eta(\tau)^3}\right|^2\,
\nn
&& \hspace{3cm}
\times
\sum_{n,w\in \bz}\,\int _0^{\infty}dp\, 
q^{\frac{p^2}{2} + \frac{1}{4}\left(\frac{n}{\sqrt{k}} + \sqrt{k} w\right)^2}
\overline{
q^{\frac{p^2}{2} + \frac{1}{4}\left(\frac{n}{\sqrt{k}} - \sqrt{k} w\right)^2}
}\, y^{\frac{n+kw}{k}} \overline{y^{-\frac{n-kw}{k}}}
\nn
&& \hspace{6cm}
+ \Z_{\msc{finite}}(\tau, z, \bar{z}) + O(\ep, \ep\ln \ep),
\label{asp part}
\end{eqnarray}
where $\cC$ is some positive constant independent of $\ep$. 
The leading part proportional to $\left| \ln \ep \right|$
(the `asymptotic part')
is  expressible in terms of the extended continuous characters 
\cite{ES-L,ES-BH} as in the free theory 
when the level $k$ is rational. 
We here denote the term of order $O(\ep^0)$ as  
$\Z_{\msc{finite}}(\tau, z, \bar{z})$, 
which is still finite after taking the $\ep\, \rightarrow \, +0$ limit and is 
modular invariant. This part can be directly extracted from 
$\Z_{\msc{reg}}(\tau,z,\bar{z} ; \ep)$ as in \cite{ncpart-NENS5};
\begin{equation}
\Z_{\msc{finite}}(\tau, z, \bar{z}) = \lim_{\ep\,\rightarrow\, +0}\,
\left[1- \ep \ln \ep \frac{\del}{\del \ep} \right] \Z_{\msc{reg}}(\tau,z,\bar{z} ; \ep).
\label{finite part formula}
\end{equation}
We emphasize that $\Z_{\msc{finite}}(\tau, z, \bar{z})$ is uniquely determined 
irrespective of the adopted regularization scheme, even though the asymptotic part 
(the overall constant $\cC$) as well as the correction terms of $O(\ep, \ep \ln \ep)$
will depend on the method of regularization.


The elliptic genus is obtained by formally setting $\bar{z}=0$ while fixing $z$ 
at a general complex value in $\Z_{\msc{finite}}(\tau, z, \bar{z})$. 
We also have to divide the partition function by a factor
\begin{equation}
e^{2\pi \frac{\hc}{\tau_2} \left( |z|^2 - z_2^2\right)  } \equiv e^{2\pi  \frac{\hc}{\tau_2} \left(\frac{z+\bar{z}}{2}\right)^2} 
\sim e^{\frac{\pi}{\tau_2} \frac{\hc}{2}z^2},
\label{anomaly factor 1}
\end{equation}
in order to include the anomaly factor correctly.
In fact, we can uniquely determine this factor by requiring the following conditions due to the analysis given in \cite{ES-NH};
\begin{itemize}
\item The elliptic genus should have the correct modular property;
\begin{eqnarray}
&& \cZ(\tau+1,z) = \cZ(\tau,z), \hspace{1cm}
\cZ\left(-\frac{1}{\tau}, \frac{z}{\tau} \right) 
= e^{i\pi \frac{\hc}{\tau} z^2}\, 
\cZ(\tau,z),
\label{modular EG}
\end{eqnarray}

\item The elliptic genus should be expanded by the discrete characters \eqn{chd u} (with $u=0$) 
around  $\tau \sim i\infty$ as 
\begin{eqnarray}
\cZ(\tau,z) & = &  \sum\, \chd(*,0;\tau,z) + [\mbox{subleading terms}],
\label{IR part EG}
\end{eqnarray}
{\em with no extra overall 
factor\/},  
where the `subleading terms' include 
spectral flow sectors as well as non-holomorphic corrections. 

\end{itemize}

In this way, 
the elliptic genus is written as 
\begin{eqnarray}
\cZ(\tau,z) & = & e^{-\frac{\pi}{\tau_2} \frac{\hc}{2}z^2} \,  
\Z_{\msc{finite}}(\tau,z,\bar{z}=0)
\nn
& \equiv & \lim_{\ep \rightarrow +0}\, 
e^{-\frac{\pi}{\tau_2} \frac{\hc}{2}z^2} \,  
\Z_{\msc{reg}}(\tau,z,\bar{z}=0 ; \ep).
\label{EG path int 0}
\end{eqnarray}
The equality of second line is due to the simple fact that 
the asymptotic term in \eqn{asp part} drops off when setting $\bar{z}=0$.
We thus obtain the path-integral expression of elliptic genus as 
\begin{eqnarray}
\hspace{-5mm}
\cZ(\tau,z) &=& 
\lim_{\ep \rightarrow +0}\, 
k e^{\frac{\pi z^2}{k\tau_2}} \, \int_{\bc} \frac{d^2\mu}{\tau_2}\, 
\sigma(\tau, \mu,z,0; \ep)\,
\frac{\th_1\left(\tau, \mu + \frac{k+2}{k}z\right)}
{\th_1\left(\tau, \mu + \frac{2}{k} z \right)} \, 
e^{2\pi i z \frac{\mu_2}{\tau_2}}\,
e^{-\frac{\pi k}{\tau_2}\left| \mu \right|^2}.
\label{EG path int}
\end{eqnarray}
The $\mu$-integral is not  easy to perform since 
the Gaussian factor 
$e^{-\frac{\pi k}{\tau_2} \left| \mu \right|^2}$ 
breaks the periodicity of the integrand.
If this factor was absent, the relevant integral would reduce to 
a simple period integral over a torus $\Sigma \equiv \bc/\La$, ($\La \equiv \bz\tau+\bz$).

One way to avoid this complication is given by using 
the following identity \cite{orb-ncpart} 
\begin{equation}
\cZ(\tau,z) = \sum_{\la \equiv n_1\tau + n_2 \in \La}\,
(-1)^{n_1+n_2 + n_1 n_2}\,  
s^{(\frac{\hc}{2})}_{\la} \cdot \cZ^{(\infty)}(\tau,z).
\label{Z sflow formula}
\end{equation}
where $s^{(\kappa)}_{\la}$ denotes the spectral flow (Eichler-Zagier \cite{Eich-Zagier}) operator defined by
\begin{eqnarray}
s^{(\kappa)}_{\la} \cdot f(\tau,z) &:= &
q^{\kappa \al^2} y^{2\kappa \al} e^{2\pi i \kappa \al \beta}\,
f(\tau,z+\la)
\nn
& \equiv & e^{2\pi i \frac{\kappa}{\tau_2} \la_2 \left(\la + 2z\right) }\,
f(\tau, z+\la),
\nn
&& \hspace{1cm} 
(\la \equiv \al \tau +\beta\equiv \lambda_1+i\lambda_2, ~ \al, \beta, \lambda_1,\lambda_2 \in \br).
\label{sflow op}
\end{eqnarray}

Recall that the elliptic genus of a complex $D$-dimensional manifold is a Jacobi form with index ${D\over 2}$ \cite{Eich-Zagier}. Here $\hat{c}$ (\ref{chat}) is the effective dimension of a target manifold described by a superconformal field theory with a central charge $c$. Thus the suffix ${\hat{c} \over 2}$ of the flow operator $s_{\lambda}^{({\hat{c}\over 2})}$ denotes the index of the 
elliptic genus $\cZ(\tau,z)$ describing the cigar geometry.

Since \eqn{Z sflow formula}
is a crucial step in our analysis, we will present 
its proof in Appendix B.
What is important here is the fact that the action of the operator $s^{(\kappa)}_{\la}$ 
(as well as the sign factor $(-1)^{n_1+n_2+n_1n_2}$) 
preserves the modular covariance, whose precise meaning
is given in Appendix A.

On the other hand, $\cZ^{(\infty)}(\tau,z)$ is defined schematically as 
the elliptic genus of `$\bz_{\infty}$-orbifold' of the cigar model as given in 
\eqn{def EG infty u}, 
or equivalently the universal cover of trumpet in the $T$-dual picture. 
Namely, 
\begin{eqnarray}
\hspace{-1cm}
\cZ^{(\infty)}(\tau,z) 
&: =& 
\lim_{\ep \rightarrow +0}\, 
k e^{\frac{\pi z^2}{k\tau_2}} \, 
\int_{\Sigma} \frac{d^2 \om}{\tau_2} \,
\int_{\bc} \frac{d^2 \mu}{\tau_2}\,
\sigma(\tau, \mu ,z,0; \ep)\,
\frac{\th_1\left(\tau, \mu + \frac{k+2}{k}z\right)}
{\th_1\left(\tau, \mu  + \frac{2}{k} z \right)} 
\, e^{2\pi i z \frac{\mu_2}{\tau_2}}\,
e^{-\frac{\pi k}{\tau_2}\left| \mu+ \om \right|^2}
\nn
&=& 
\lim_{\ep \rightarrow +0}\, 
 e^{\frac{\pi z^2}{k\tau_2}} \, 
\int_{\Sigma} \frac{d^2 \mu}{\tau_2}\, \sigma(\tau, \mu ,z,0; \ep)\,
\frac{\th_1\left(\tau, \mu + \frac{k+2}{k}z\right)}
{\th_1\left(\tau, \mu  + \frac{2}{k} z \right)} 
\, e^{2\pi i z \frac{\mu_2}{\tau_2}}.
\label{eval EG infty}
\end{eqnarray}
In the second line, 
due to the periodicity of the integrand
under $\mu\, \rightarrow \, \mu + \nu$, $\nu \in \La$
 except for the factor 
$
e^{-\frac{\pi k}{\tau_2}\left| \mu+ \om \right|^2}
$, 
we made use of an obvious relation 
$
\int_{\Sigma} \frac{d^2 \om}{\tau_2} \, 
\int_{\bc} \frac{d^2 \mu}{\tau_2}\, \left[\cdots \right]
= \int_{\bc} \frac{d^2 \om}{\tau_2} \, 
\int_{\Sigma} \frac{d^2 \mu}{\tau_2}\, \left[\cdots \right],
$
and  carried out the Gaussian integral over $\om$.

Computation is now easy.
Since the integrand of \eqn{eval EG infty}  
is holomorphic and periodic with respect to the integration
variables $s_1$, $s_2$ with 
$\mu \equiv s_1\tau+s_2$, 
one may regard it as a double period integral.
Thus,  by deforming the integration contour as 
\begin{equation}
s_1 \in [0,1] + i\frac{z}{k\tau_2}, \hspace{1cm}
s_2 \in [0,1] - i\frac{z\bar{\tau}}{k\tau_2}, 
\label{contour deform 1}
\end{equation}
we can directly evaluate \eqn{eval EG infty}  
with the helps of the identity \eqn{theta id} as 
\begin{eqnarray}
\cZ^{(\infty)}(\tau,z) & = & \lim_{\ep\, \rightarrow \, +0} 
e^{- \frac{\pi z^2}{k\tau_2}}\, 
\int_{\Sigma} \frac{d^2 \mu}{\tau_2}\, 
\sigma(\tau, \mu,0, 0;\ep)
\frac{\th_1\left(\tau, \mu + z\right)}
{\th_1\left(\tau, \mu  \right)} \, 
\, e^{2\pi i z \frac{\mu_2}{\tau_2}}
\nn
&=& 
e^{- \frac{\pi z^2}{k\tau_2}}\, 
\int_{\Sigma} \frac{d^2 \mu}{\tau_2}\, 
\frac{\th_1\left(\tau, \mu + z\right)}
{\th_1\left(\tau, \mu  \right)} \, 
\, e^{2\pi i z \frac{\mu_2}{\tau_2}}
\nn
&=& e^{- \frac{\pi z^2}{k\tau_2}}\, 
\frac{\th_1(\tau,z)}{i\eta(\tau)^3}\, \sum_{n\in \bz}\, \int_{\Sigma} \frac{d^2 \mu}{\tau_2}\, 
\frac{e^{2\pi i n\mu }}{1-yq^n}\, e^{2\pi i z \frac{\mu_2}{\tau_2}}
\nn
&=& e^{- \frac{ \pi z^2}{k\tau_2}}\,  
\frac{\th_1(\tau,z)}{2\pi z \eta(\tau)^3}
\equiv 
e^{- \frac{ \pi z^2}{k\tau_2}}\frac{\vth(\tau,z)}{z} .
\label{EG infty}
\end{eqnarray}
In the second line, we have used  
$$
\sigma(\tau, \mu,0,0 ;\ep) = 
 \prod_{m_1,m_2 \in \bz}\, 
\left[ 1- e^{- \frac{1}{\ep \tau_2} \left|\mu + m_1 \tau+m_2\right|^2} \right],
$$
as well as the fact that the $\mu$-integral converges even in the absence of 
$\sigma$-factor.  
We also introduced the symbol \eqn{def vth};
$$
\vth(\tau,z) \equiv \frac{\th_1(\tau,z)}{2\pi \eta(\tau)^3}.
$$

Substituting \eqn{EG infty} back into the `spectral flow formula'  
\eqn{Z sflow formula},  
we finally obtain the following simple expression of elliptic genus of cigar model;
\begin{eqnarray}
\cZ(\tau,z) & = & 
\sum_{\la \equiv n_1\tau + n_2 \in \La}\,
(-1)^{n_1+n_2 + n_1 n_2}\,  
s^{(\frac{\hc}{2})}_{\la} \cdot
\left[
e^{- \frac{ \pi z^2}{k\tau_2}}\,  
\frac{\vth(\tau,z)}{z}
\right]
\nn
& =& 
\vth(\tau,z) \, 
\sum_{\la \in \La}\,
s^{(\frac{1}{k})}_{\la} \cdot
\left[
\frac{e^{- \frac{ \pi z^2}{k\tau_2}}}{z}
\right]
\nn
& = &  
\vth(\tau,z) \, 
\sum_{\la \in \La}\, 
\frac{
e^{- \frac{\pi}{k\tau_2} \left[ z^2+  |\la|^2 + 2 \bar{\la} z \right] }
}
{z+\la}
\equiv 
\vth(\tau,z) \, 
\sum_{\la \in \La}\, 
\frac{
\rho^{(1/k)}(\la,z)
}
{z+\la}.
\label{EG nh-E}
\end{eqnarray}
In the last line, we introduced the notation 
\begin{equation}
\rho^{(\kappa)}(\la,z) := s^{(\kappa)}_{\la} \cdot 
e^{-\frac{\pi \kappa}{\tau_2} z^2} \equiv 
e^{- \frac{\pi \kappa}{\tau_2} \left[|\la|^2 + 2\bar{\la} z + z^2\right] }.
\label{def rho}
\end{equation}
More explicitly, \eqn{EG nh-E} is rewritten like a Poincar\'e series
\begin{eqnarray}
\cZ(\tau,z) = \vth(\tau,z) \, 
\sum_{m, n\in \bz}\, q^{\frac{1}{k} m^2} y^{\frac{2}{k} m} e^{2\pi i \frac{mn}{k}}\, 
\frac{ e^{-\frac{\pi }{k \tau_2} (z+m\tau+n)^2}}{z+m\tau+n}.
\label{EG nh-E 2}
\end{eqnarray}

Here the double power series of $\la \in \La$
absolutely converges due to the Gaussian factor, 
and thus \eqn{EG nh-E} exhibits good modular behavior.
One can easily confirm that the elliptic genus \eqn{EG nh-E} 
possesses the modular property as a weak Jacobi form with weight 0 and index $\hat{c}/2$ given in \eqn{modular EG},
and also for the case of $k= N/K$, $N, K \in \bz_{>0}$, 
\begin{eqnarray}
&& s^{(\frac{\hc}{2})}_{n_1\tau+n_2} \cdot \cZ(\tau,z) = (-1)^{n_1+n_2 + n_1 n_2}
 \cZ(\tau,z), \hspace{1cm}
 (\any n_1, n_2 \in N \bz).
 \label{sflow EG}
\end{eqnarray}


~


\subsection{Relations Between the New and Old Expressions of Modular 
Completions}

Now, let us discuss on the relations between present  results 
and our previous analyses given in \cite{ES-NH,orb-ncpart},
which were based on the `character decomposition' of the torus partition function.

First of all, we would like to point out that 
$\cZ^{(\infty)}(\tau,z)$ is expanded in terms of modular completions of 
the {\em irreducible\/} discrete characters, 
\begin{equation}
 \cZ^{(\infty)} (\tau,z) = \frac{1}{k} \int_0^k d\nu \, \hchd(\nu,0;\tau,z),
\label{rel EG infty irred mod comp} 
\end{equation}
as was shown in \cite{orb-ncpart}.
Here, the modular completion $\hchd(\nu, n;\tau,z)$ ($0\leq \nu \leq k$, $n\in \bz$)
has been defined by \eqn{hchd u}.
It would be helpful to confirm the consistency of the formulas \eqn{EG infty}
and \eqn{rel EG infty irred mod comp}.  
In fact, by using \eqn{hchd u}, the R.H.S of \eqn{rel EG infty irred mod comp}
is explicitly written as 
\begin{eqnarray}
\frac{1}{k} \int_0^k d\nu \, \hchd(\nu,0;\tau,z) &=& 
\frac{1}{k} \frac{\th_1(\tau,z)}{i\eta(\tau)^3} \, 
\left[ \int_{0}^{k} d\nu \, \frac{y^{\frac{\nu}{k}}}{1-y} 
+ \frac{i}{2\pi} \int_{-\infty}^{\infty}d\nu \int_{-\infty}^{\infty} dp\, 
\frac{y^{\frac{\nu}{k}} e^{-\pi \tau_2 \frac{p^2+\nu^2}{k}}}{p-i \nu}\,
\right].
\nn
&&
\label{rel EG infty irred mod comp 2}
\end{eqnarray}
The first term corresponds to the holomorphic part, which just yields
\begin{equation}
 \frac{1}{k} \int_0^k d\nu \, \chd(\nu,0;\tau,z) \equiv 
 \frac{1}{k} \frac{\th_1(\tau,z)}{i\eta(\tau)^3} \, 
\int_{0}^{k} d\nu \, \frac{y^{\frac{\nu}{k}}}{1-y} = 
\frac{\vth(\tau,z)}{z}.
\label{EG infty hol}
\end{equation}
On the other hand, the double integral in the second term can 
be evaluated by using the polar coordinate
$
r e^{i\theta} \equiv p+i\nu 
$
and elementary properties of Bessel function
as follows;
\begin{eqnarray}
\frac{1}{k} \int_{-\infty}^{\infty} d\nu \, \int_{-\infty}^{\infty} dp\,
\frac{  y^{\frac{\nu}{k}}e^{-\pi \tau_2 \frac{p^2+\nu^2}{k}}}{p-i\nu}
&=& \frac{1}{k} \int_0^{\infty} r dr\, \int_0^{2\pi} d\theta\, 
\frac{e^{-\pi \tau_2 \frac{r^2}{k}}}{r e^{-i\theta}}
\cdot e^{2\pi i z \frac{r}{k} \sin \theta}
\nn
& =& - \frac{2\pi}{k} \int_0^{\infty} dr\, e^{-\pi \tau_2 \frac{r^2}{k}}\, J_1 (2\pi  z r/k )
\nn
&=& \sum_{m=0}^{\infty} \frac{(-1)^{m+1}}{(m+1)!} \, 
\left(\frac{\pi }{k \tau_2}\right)^{m+1} z^{2m+1} 
\nn
&=& \frac{1}{z} \left(e^{- \frac{\pi z^2}{k\tau_2}} -1\right).
\label{EG infty nh}
\end{eqnarray}
Substituting \eqn{EG infty hol} and \eqn{EG infty nh} into 
\eqn{rel EG infty irred mod comp 2}, 
we recover the anticipated formula \eqn{EG infty};
\begin{eqnarray}
\cZ^{(\infty)} (\tau,z) &=& 
\frac{\vth(\tau,z)}{z}
+
\frac{\vth(\tau,z)}{z}
\left(e^{- \frac{\pi z^2}{k\tau_2}} -1\right)
\nn
&=& e^{- \frac{\pi z^2}{k\tau_2}}
\frac{\vth(\tau,z)}{z}.
\label{EG infty 2}
\end{eqnarray}

~


We next recall the formula of the elliptic genus of cigar model in the case of 
$k=N/K$, ($N,K \in \bz_{>0}$)
given in \cite{ES-NH};
\begin{eqnarray}
\cZ(\tau,z) & = & 
\frac{\th_1(\tau,z)}{i\eta(\tau)^3} \, \frac{1}{N} \sum_{a,b\in \bz_N}\,
s_{a\tau+b}^{(NK)} \cdot \hf^{(NK)}\left(\tau, \frac{z}{N}\right),
\label{EG cigar formula}
\end{eqnarray}
where $\hf^{(*)}(\tau,z) \equiv \hf^{(*)}_0(\tau,z)$ is the function 
given in \eqn{hAppell}.
Equating this formula with our present result \eqn{EG nh-E} in the special  case of 
$N=1$, $K=1/k \in \bz_{>0}$,   
one can rewrite $\hf^{(K)}$
in a compact form as  
\begin{eqnarray}
\hf^{(K)}(\tau,z) & =& \frac{i}{2\pi} \,
\sum_{\nu \in \La} \, 
\frac{
\rho^{(K)}(\nu,z)}
{z+\nu}
\nn
&\equiv&
\frac{i}{2\pi} \sum_{m,n \in \bz }\, q^{K m^2 } y^{2K m} \frac{e^{-\frac{\pi K}{\tau_2} (z+m\tau+n)^2 }}{z+m\tau+n}.
\label{hAppell nh-E}
\end{eqnarray}
We shall present a direct proof of this formula in Appendix C,  with the `$u$-parameter' included.

We also recall the relations \cite{ES-NH,orb-ncpart}
\begin{eqnarray}
\cZ(\tau,z)  &=&  
\sum_{\stackrel{v, a \in \bz_N}{v+Ka \in N\bz}}\, 
\hchid^{(N,K)}(v,a;\tau,z) 
\nn
& \equiv &  \sum_{\stackrel{v, a \in \bz_N}{v+Ka \in N\bz}}\, 
\sum_{m\in a + N\bz}\, \hchd\left(\frac{v}{K}, m;\tau,z\right), 
\label{EG cigar formula 2}
\end{eqnarray}
where $\hchid^{(N,K)}$ denotes the modular completion of the discrete extended character \eqn{hchid u} (for the $u=0$ case).
We then likewise obtain 
\begin{eqnarray}
\hchid^{(N,K)}(v,a;\tau,z) & = & \vth(\tau,z) \,
\sum_{b\in \bz_N}\, \sum_{\la\in a\tau+b + N\La} \,e^{-2\pi i \frac{b}{N}(v+Ka)}\,
\frac{\rho^{\left(1/k\right)}(\la,z)}{z+\la},
\label{hchid nh-E}
\\
\hchd(\nu,m;\tau,z)  &= &\vth(\tau,z) \, 
\sum_{n\in \bz}\, e^{-2\pi i \frac{n}{k}(\nu+m)}\,
\frac{\rho^{(1/k)} (m\tau+n, z)}{z+m\tau+n}.
\label{hchd nh-E}
\end{eqnarray}


~


We would like to make a comment:

~

The above manipulations leading to \eqn{EG nh-E} is easily generalized 
to the `twisted' case;
\begin{eqnarray}
\hspace{-1cm}
\cZ_{\om}(\tau,z) 
& : = & 
\lim_{\ep\rightarrow +0}\,
k \, e^{\frac{\pi}{k \tau_2} z^2} \, 
\int_{\bc} \frac{d^2 \mu}{\tau_2}\, 
\sigma(\tau,\mu,z, 0; \ep)\,
\frac{\th_1 \left(\tau, \mu +  \frac{k+2}{k}z  \right)}
{\th_1\left(\tau, \mu + \frac{2}{k}z \right)} 
\, e^{2 \pi i z \frac{\mu_2}{\tau_2}}\,
e^{-\frac{\pi k}{\tau_2}\left| \mu+ \om \right|^2},
\label{def twisted EG}
\end{eqnarray}
where $\om \in \bc/ \La ( \equiv \Sigma )$ is the twist parameter.
(We have an obvious periodicity under 
$\om\, \rightarrow \om +  \nu$, ~ $\any \nu \in \La$.)

As is mentioned in Appendix B, 
the formula of `spectral flow decomposition'  
\eqn{Z sflow formula} is generalized  to the present case, which is 
naturally  interpreted as the  Fourier expansion written as 
\begin{eqnarray}
&& \cZ_{\om}(\tau,z) = \sum_{\nu \in \La}\, e^{2\pi i \frac{1}{\tau_2} 
\Im (\om \bar{\nu})}\, \tcZ^{(\infty)}_{\nu} (\tau,z),
\label{twisted EG Fourier rel}
\end{eqnarray}
Here, we introduced 
\begin{eqnarray}
\tcZ_{\nu}^{(\infty)}(\tau,z) &:=& (-1)^{n_1+n_2+n_1n_2}\, s^{(\frac{\hc}{2})}_{\nu}
\cdot  \cZ^{(\infty)}(\tau,z)
\nn
& \equiv & \vth(\tau,z) \, \frac{\rho^{(\frac{1}{k})} (\nu,z)}{z+\nu},
\hspace{1cm} (\nu\equiv n_1 \tau+ n_2 \in \La),
\label{tcZ nu}
\end{eqnarray} 
which is also expressed in terms of the modular completion $\hchd(\la, n)$ \eqn{hchd u}
in the same way as 
\eqn{rel EG infty irred mod comp};
\begin{equation}
\tcZ_{\nu}^{(\infty)}(\tau,z) = \frac{1}{k} \int_0^k d\la \, e^{2\pi i \frac{n_2}{k}(\la + n_1)} \, \hchd (\la, n_1 ; \tau,z).
\label{rel EG infty twisted irred mod comp}
\end{equation}

The identity \eqn{twisted EG Fourier rel} looks nice. This schematically 
means that 
 the spectral flow with $\nu \in \La$ is the Fourier transform of the
twisting by $\om \in \bc/\La$. 


Each manipulation given here keeps the modularity intact. 
We can readily confirm 
\begin{eqnarray}
&& \hspace{-5mm}
\cZ_{\om} (\tau+1,z) = \cZ_{\om} (\tau,z), \hspace{1cm}
\cZ_{\frac{\om}{\tau}}\left(-\frac{1}{\tau}, \frac{z}{\tau} \right) 
= e^{i\pi \frac{\hc}{\tau} z^2}\, 
\cZ_{\om}(\tau,z),
\label{modular twisted EG 1}
\\
&& \hspace{-5mm}
\tcZ^{(\infty)}_{\nu} (\tau+1,z) = \tcZ^{(\infty)}_{\nu} (\tau,z), \hspace{1cm}
\tcZ^{(\infty)}_{\frac{\nu}{\tau}}\left(-\frac{1}{\tau}, \frac{z}{\tau} \right) 
= e^{i\pi \frac{\hc}{\tau} z^2}\, 
\tcZ^{(\infty)}_{\nu}(\tau,z).
\label{modular twisted EG 2}
\end{eqnarray}


One can also  evaluate the elliptic genus of 
the $\bz_M$-orbifold of cigar with $\any M\in \bz_{>0}$ as
\begin{eqnarray}
\cZ^{(M)} (\tau,z) & \equiv  & \frac{1}{M} \sum_{\gamma \in \La/M\La}\,
\cZ_{\frac{\gamma}{M}} (\tau,z) 
\nn
& =& \frac{1}{M} \sum_{\nu \in \La}\,  \sum_{\gamma \in \La/M\La}\,
e^{2\pi i \frac{1}{M\tau_2} \Im (\gamma \bar{\nu} )} \, \tcZ^{(\infty)}_{\nu} (\tau,z)
\hspace{1cm} 
\left(\because ~ \eqn{twisted EG Fourier rel}\right)
\nn
&=& M \sum_{\nu \in M \La}\, \tcZ^{(\infty)}_{\nu} (\tau,z)
\equiv  M \vth(\tau,z) \, \sum_{\nu \in M \La}\, 
\frac{\rho^{(1/k)}(\nu ,z)}{z+\nu}.
\label{Z M nh-E-series}
\end{eqnarray}
More generally, 
rewriting $\cZ^{(M)}_{\gamma} (\tau,z) \equiv \cZ_{\gamma/M} (\tau,z)$ as in Appendix B, 
the Fourier transformation is calculated as 
($\any \la \equiv m_1 \tau+ m_2 \in \La/M\La$);
\begin{eqnarray}
\tcZ^{(M)}_{\la} (\tau,z) & \equiv  & \frac{1}{M}  \sum_{\gamma \in \La/M\La}\,
e^{2\pi i \frac{1}{M\tau_2} \Im (\la \bar{\gamma})}
\,\cZ^{(M)}_{\gamma} (\tau,z) 
\nn
\nn
&=& M \vth(\tau,z) \, \sum_{\nu \in \la +M\La}\, \frac{\rho^{(1/k)}(\nu ,z)}{z+\nu}
\nn
& = &  (-1)^{m_1+m_2+m_1m_2} s^{(\frac{\hc}{2})}_{\la} \cdot \cZ^{(M)}(\tau,z).
\label{Fourier relation general}
\end{eqnarray}
This result is consistent with the identity \eqn{Fourier relation 2}, which is directly 
checked based on the path-integral expression \eqn{def twisted EG} 
for the special cases  $k=N/K$, $M=N$, as is mentioned in Appendix B. 
It would be worthwhile to note that \eqn{Fourier relation general} is always correct 
for $\any M \in \bz$ and $\any k \in \br_{>0}$ (not necessarily rational).



~

~


\section{Extension to General Spin Structures}

\subsection{Elliptic Genus with General Spin Structures}

In this section, we shall extend the analyses in the previous section to 
the cases of general spin structure with the {\em continuous parameter\/} of twisting 
$u\equiv \al \tau+ \beta $, $(\any \al, \beta \in \br)$.
Namely, we assume the world-sheet fermions of super-gauged WZW model $\psi^{\pm} (w)$, $\tpsi^{\pm}(\bar{w})$ 
satisfy the twisted  boundary condition (with respect to the cylinder coordinate $w$, $\bar{w}$); 
\begin{eqnarray}
&& \psi^{\pm}(w+1) = e^{\mp 2\pi i \al } \psi^{\pm}(w), \hspace{1cm} 
\psi^{\pm} ( w+ 2\pi \tau) = e^{\mp 2\pi i \beta} \psi(w),
\nn
&& \tpsi^{\pm} (\bar{w}+1) = e^{ \pm 2\pi i \al } \tpsi^{\pm}(\bar{w}), \hspace{1cm} 
\tpsi^{\pm} ( \bar{w}+ 2\pi \bar{\tau}) = e^{\pm 2\pi i \beta} \tpsi^{\pm}(\bar{w}).
\label{twisted bc fermion}
\end{eqnarray}
In the end, this twist parameter $u$ will turn out to be identified with the `$u$-variable' of the function $\hf^{(*)}_u(\tau,z)$ \label{hAppell u} 
introduced in \cite{Zwegers}.
Closely related studies including such a twisting based on a different approach 
have been given in \cite{AT,AT3}. 

We shall again start our analysis from the torus partition function. 
The partition function for the fermions with boundary condition \eqn{twisted bc fermion} 
is written concisely  by using the spectral flow operator $s^{(1/2)}_{-u}$ as  
\begin{equation}
\bm{Z}_{\msc{fermion}}(\tau, z, u) = e^{-\frac{2\pi}{\tau_2} z_2^2}\,
\left|\frac{s^{(1/2)}_{-u}\cdot \th_1(\tau,z)}{\eta(\tau)}\right|^2,
\end{equation}
which is manifestly modular invariant. 
Thus, the regularized partition function \eqn{Z reg} is now replaced by  
\begin{eqnarray}
\Z_{\msc{reg}}(\tau,z, \bar{z}, u ; \ep) &=&  
k e^{2\pi \frac{\hc}{\tau_2} |z|^2 - 2\pi \frac{k+4}{k\tau_2} z_2^2} \, 
\int_{\bc} \frac{d^2 \mu}{\tau_2}\, 
\sigma(\tau, \mu,z, \bar{z}; \ep)\,
\nn
&& 
\hspace{2cm}
\times
\left|
\frac{s_{-u}^{(1/2)}\cdot \th_1 \left(\tau, \mu + \frac{k+2}{k}z\right)}
{\th_1\left(\tau, \mu + \frac{2}{k} z \right)}\right|^2 
\, e^{- 4\pi  z_2 \frac{\mu_2}{\tau_2}}\,
e^{-\frac{\pi k}{\tau_2}\left| \mu \right|^2}
\nn
& \equiv  & 
k e^{2\pi \frac{\hc}{\tau_2} |z|^2 - \frac{2\pi}{k\tau_2} \left\{
(k+4) X_2^2 +  4 X_2 u_2\right\}} \, 
\int_{\bc} \frac{d^2 \mu}{\tau_2}\, 
\sigma(\tau, \mu,X+u, \bar{X}+\bar{u}; \ep)\,
\nn
&& \hspace{2cm}
\times
\left|
\frac{\th_1 \left(\tau, \mu + \frac{k+2}{k}X + \frac{2}{k} u\right)}
{\th_1\left(\tau, \mu + \frac{2}{k} (X+u) \right)}\right|^2 
\, e^{- 4\pi  X_2 \frac{\mu_2}{\tau_2}}\,
e^{-\frac{\pi k}{\tau_2}\left| \mu \right|^2}.
\label{Z reg u}
\end{eqnarray}
In the second line we set $X:= z-u$ for convenience of calculation given below. 
Note that we use the notations 
$z\equiv z_1+iz_2, ~ u\equiv u_1+iu_2,~ X\equiv X_1+iX_2,~ \mu\equiv \mu_1+i\mu_2$, with  $ z_i, u_i, X_i, \mu_i \in {\br}$, for $i=1,2$.


The elliptic genus is again defined by setting $\bar{z}$ to a particular value  while  
keeping $z$ at a generic complex value. 
It is easy to find that we have to set $\bar{z}=\bar{u}$ ($\Longleftrightarrow~ \bar{X} = 0$)
to realize the supersymmetric cancellation. 
We still have to divide by a suitable anomaly factor corresponding to 
\eqn{anomaly factor 1} in the $u=0$ case. 
To determine this factor, we require the following conditions
which extend \eqn{modular EG} and \eqn{IR part EG} 
to the $u\neq 0$ cases:
\begin{itemize}
\item The elliptic genus should have the modular property;
\begin{eqnarray}
&& \cZ(\tau+1,z,u) = \cZ(\tau,z,u), \hspace{1cm}
\cZ\left(-\frac{1}{\tau}, \frac{z}{\tau}, \frac{u}{\tau} \right) 
= e^{i\pi \frac{\hc}{\tau} z^2}\, 
\cZ(\tau,z,u),
\label{modular EG u}
\end{eqnarray}

\item The elliptic genus should be expanded around  
$\tau \sim i\infty$ in the form  as
\begin{eqnarray}
\cZ(\tau,z,u) & = &  \sum\, \chd(*,0;\tau,z,u) + [\mbox{subleading terms}]
\nn
& \equiv & \sum\, s^{(\frac{\hc}{2})}_{-u}\cdot \chd(*,0;\tau,z) 
+ [\mbox{subleading terms}].
\label{cond IR part EG u}
\end{eqnarray}

\end{itemize}
Then the wanted anomaly factor turns out to be
\begin{equation}
e^{2\pi \frac{\hc}{\tau_2} \left(z_1^2 + i\frac{u_2}{2} \bar{u} \right)} 
\sim e^{ 2\pi \frac{\hc}{\tau_2} \left\{ \left(\frac{z+\bar{u}}{2}\right)^2  + i \frac{u_2}{2} \bar{u} \right\} }
\equiv \rho^{(\frac{\hc}{2} )}(u,z)^{-1}.
\label{anomaly factor u} 
\end{equation}
Namely, we define the elliptic genus as
\begin{eqnarray}
\cZ(\tau,z,u) & : = & \rho^{(\frac{\hc}{2} )}(u,z) \, \lim_{\ep\rightarrow +0}\, 
\Z_{\msc{reg}}(\tau,z, \bar{z}=\bar{u}, u ; \ep)
\nn
& = & e^{-\frac{\pi}{k\tau_2} \left(u^2 -|u|^2\right)} e^{i\pi \frac{u_2}{\tau_2} \left(u-2z\right) }\, \lim_{\ep\rightarrow +0}\,
k \, e^{\frac{\pi}{k \tau_2} X^2} \, 
\int_{\bc} \frac{d^2 \mu}{\tau_2}\, 
\sigma(\tau, \mu,X+u, \bar{u}; \ep)\,
\nn
&& \hspace{2cm}
\times
\frac{\th_1 \left(\tau, \mu + X + \frac{2}{k}(X+ u) \right)}
{\th_1\left(\tau, \mu + \frac{2}{k} (X+u) \right)} 
\, e^{2 \pi i X \frac{\mu_2}{\tau_2}}\,
e^{-\frac{\pi k}{\tau_2}\left| \mu \right|^2}.
\label{EG cigar u}
\end{eqnarray}
Perhaps, the simplest way to compute this integral is again to use  
the `spectral flow method' that generalizes \eqn{Z sflow formula}; 
\begin{equation}
\cZ(\tau,z, u) = \sum_{\la \equiv n_1\tau + n_2 \in \La}\,
(-1)^{n_1+n_2 + n_1 n_2}\,  e^{\frac{2 \pi i}{\tau_2} \Im (\bar{\la}u)} \,
s^{(\frac{\hc}{2})}_{\la} \cdot \cZ^{(\infty)}(\tau,z,u).
\label{Z sflow formula u}
\end{equation}
We will give a proof of this identity in Appendix B.

$\cZ^{(\infty)}(\tau,z,u)$ is 
evaluated in the same way as \eqn{eval EG infty} and \eqn{EG infty} for 
the $u=0$ case ;
\begin{eqnarray}
\hspace{-0.5cm}
\cZ^{(\infty)}(\tau,z, u) 
&: =& 
e^{-\frac{\pi}{k\tau_2} \left(u^2 -|u|^2\right)} e^{i\pi \frac{u_2}{\tau_2} \left(u-2z\right) }\,
\lim_{\ep \rightarrow +0}\, 
k e^{\frac{\pi X^2}{k\tau_2}} \, \int_{\Sigma} \frac{d^2 \om}{\tau_2}\,
\int_{\bc} \frac{d^2 \mu}{\tau_2} \,  
\sigma(\tau, \mu ,X+u,\bar{u}; \ep)\,
\nn 
&& 
\hspace{2cm}
\times 
\frac{\th_1\left(\tau, \mu + X+ \frac{2}{k}(X+u)\right)}
{\th_1\left(\tau, \mu  + \frac{2}{k} (X+u)  \right)} 
\, e^{2\pi i X \frac{\mu_2}{\tau_2}}\, 
e^{-\frac{\pi k}{\tau_2}\left| \mu+ \om \right|^2}
\nn
&=& 
e^{-\frac{\pi}{k\tau_2} \left(u^2 -|u|^2\right)} e^{i\pi \frac{u_2}{\tau_2} \left(u-2z\right) }\,
e^{-\frac{\pi}{k\tau_2} \left\{ X^2 + 2 (u-\bar{u}) X \right\}}\,
\int_{\Sigma} \frac{d^2 \mu}{\tau_2} \frac{\th_1(\tau, \mu +X)}{\th_1(\tau,\mu)} e^{2\pi i X \frac{\mu_2}{\tau_2}}
\nn
& =& e^{-\frac{\pi}{k\tau_2} \left(u^2 -|u|^2\right)} e^{i\pi \frac{u_2}{\tau_2} \left(u-2z\right) }\,
e^{-\frac{\pi}{k\tau_2} \left\{ (z-u)^2 + 2 (u-\bar{u}) (z-u) \right\}}\, 
\frac{\vth(\tau,z-u)}{z-u}
\nn
& = & 
\vth_u(\tau,z) \frac{e^{-\frac{\pi}{k \tau_2} \left[ z^2 - 2\bar{u} z + \left|u\right|^2  \right]}}{z-u}  
\equiv \vth_u(\tau,z) \frac{\rho^{(1/k)}(-u,z)}{z-u}.
\label{EG infty u}
\end{eqnarray}
We again made use of the rewriting ~
$
\int_{\Sigma} \frac{d^2 \om}{\tau_2}\, \int_{\bc} \frac{d^2 \mu}{\tau_2} \,
\left[\cdots \right]
= \int_{\bc} \frac{d^2 \om}{\tau_2}\, \int_{\Sigma} \frac{d^2 \mu}{\tau_2} \, 
\left[\cdots \right]
$
and carried out the Gaussian integral
over $\om$.  In the third line,  we again removed the regularization factor $\sigma(*;\ep)$ by a suitable contour  deformation,  and introduced the symbol
\begin{equation}
\vth_u(\tau,z) := s^{(1/2)}_{-u}\cdot \vth(\tau,z) \equiv e^{i\pi \frac{u_2}{\tau_2} \left(u-2z\right) }\, \vth(\tau,z-u),
\label{vth u}
\end{equation}
in the last line.

We here note 
\begin{equation}
\cZ^{(\infty)}(\tau,z,u) 
= s^{(\frac{\hc}{2})}_{-u} \cdot \cZ^{(\infty)} (\tau,z)
\equiv s^{(\frac{\hc}{2})}_{-u} \cdot 
\left[\vth(\tau,z) \frac{e^{-\frac{\pi z^2}{k \tau_2}}}{z}\right].
\label{EG infty u 2}
\end{equation}
Therefore, substituting \eqn{EG infty u 2} into \eqn{Z sflow formula u},
we finally obtain
\begin{eqnarray}
\cZ(\tau,z, u) &=& \sum_{\la \equiv n_1\tau + n_2 \in \La}\,
(-1)^{n_1+n_2 + n_1 n_2}\,  e^{- \frac{2 \pi i}{\tau_2} \Im(\la \bar{u})} \,
s^{(\frac{\hc}{2})}_{\la} \cdot s^{\left(\frac{\hc}{2} \right)}_{-u} \cdot \left[\vth(\tau,z) \frac{e^{-\frac{\pi z^2}{k \tau_2}}}{z}\right]
\nn
& = & \vth_u(\tau,z) \sum_{\la  \in \La}\, 
s^{(\frac{1}{k})}_{\la} \cdot s^{\left(\frac{1}{k} \right)}_{-u} \cdot \left[ \frac{e^{-\frac{\pi z^2}{k \tau_2}}}{z}\right] \nn
& = & \vth_u(\tau,z) \sum_{\la   \in \La}\, 
e^{2\pi i \frac{1}{k\tau_2} \Im (\la \bar{u})} \,
s^{\left(\frac{1}{k} \right)}_{\la-u} \cdot \left[ \frac{e^{-\frac{\pi z^2}{k \tau_2}}}{z}\right] \nn
& = & \vth_u(\tau,z) \sum_{\la   \in \La -u}\, 
\frac{\rho^{(1/k)}(\la,z) \, e^{2\pi i \frac{1}{k\tau_2} \Im (\la \bar{u})} }{z+\la}.
\label{EG nh-E u}
\end{eqnarray}
In the second line we used the identity 
$$
s^{(1/2)}_{n_1\tau+n_2} \cdot \vth_u(\tau,z) 
= (-1)^{n_1+n_2+n_1n_2} e^{2 \pi i \frac{1}{\tau_2}
\Im (\la \bar{u})} 
\, \vth_u(\tau,z),
$$
which is equivalent with \eqn{sflow vth u},
and we also made use of the product relation of the spectral flow operators 
\eqn{product sflow op} in the third line.

The main result \eqn{EG nh-E u} is an extension of   
the formula \eqn{EG nh-E}. 
This obviously shows that $\cZ(\tau,z,u)$ possesses the expected 
modularity  \eqn{modular EG u},
and also the spectral flow property for the case of $k=N/K$, $N,K \in \bz_{>0}$
expressed as 
\begin{eqnarray}
&& s^{(\frac{\hc}{2})}_{n_1\tau+n_2} \cdot \cZ(\tau,z,u) = (-1)^{n_1+n_2 + n_1 n_2}
 \cZ(\tau,z,u), \hspace{1cm}
 (\any n_1, n_2 \in N \bz).
 \label{sflow EG u}
\end{eqnarray}


One can also write down the  extension of the formulas
\eqn{EG cigar formula} and \eqn{EG cigar formula 2};
\begin{eqnarray}
\cZ(\tau,z,u) & = & \sum_{\stackrel{v, a \in \bz_N}{v+Ka \in N\bz}}\, 
\hchid^{(N,K)}(v,a;\tau,z,u) 
\nn
& \equiv & 
-2\pi i \vth_u(\tau,z) \, e^{- \frac{\pi}{\tau_2} NK \left(u^2 -|u|^2 \right)} \,
\frac{1}{N} \sum_{a,b\in \bz_N}\,
s_{a\tau+b}^{(NK)} \cdot \hf_u^{(NK)}\left(\tau, \frac{z}{N}\right),
\label{EG cigar formula u}
\end{eqnarray}
where the modular completion of extended character $\hchid^{(N,K)}(v,a;\tau,z,u) $
is defined in \eqn{hchid u}
and 
$\hf^{(k)}_u(\tau,z)$ is the function 
given in \eqn{hAppell}.
Comparing \eqn{EG nh-E u} with \eqn{EG cigar formula u}, 
we obtain a formula, which is the extension of  
\eqn{hAppell nh-E}; 
\begin{equation}
\hf_u^{(k)}(\tau,z) = e^{\frac{\pi k}{\tau_2} \left(u^2 -|u|^2 \right)} \, \frac{i }{2\pi} \sum_{\la \in \La-u} \, 
\frac{\rho^{(k)} (\la,z) \, e^{2\pi i \frac{ k }{\tau_2} \Im(\la \bar{u})} }{z+\la}.
\hspace{1cm} (k\in \bz_{>0})
\label{hAppell nh-E u}
\end{equation}
We present a direct proof of this identity in Appendix C.

Based on \eqn{hAppell nh-E u}, one can readily confirm the following identities for  
the function $\hf^{(k)}_u(\tau,z)$;\footnote
   {It would be useful to rewrite \eqn{hAppell nh-E u}
as 
$$
\hf^{(k)}_u(\tau,z) = \frac{i}{2\pi} \sum_{\la \in \La}\, s^{(k)}_{\la}\cdot 
\left[\frac{e^{-\frac{\pi k}{\tau_2} \left\{ (z-u)^2 + 2(u-\bar{u}) (z-u) \right\}}}{z-u}\right],
$$
in order to check the spectral flow relations \eqn{sflow hAppell u} and \eqn{sflow hAppell u 2}.
}
\begin{eqnarray}
&& \hf_u^{(k)}(\tau+1,z) = \hf_u^{(k)}(\tau,z)  , 
\hspace{1cm}
\hf^{(k)}_{\frac{u}{\tau}} \left(-\frac{1}{\tau}, \frac{z}{\tau}\right)
= \tau e^{2\pi i \frac{k}{\tau}(z^2-u^2)}\,
\hf^{(k)}_u(\tau,z).
\label{modular hAppell u}
\\
&& \hf_{u+\om}^{(k)}(\tau,z+\nu) 
= e^{-2\pi i \frac{k}{\tau_2} \left[ \nu_2  (\nu+2z) - 
\om_2 (\om + 2 u)
\right] } \, \hf_u^{(k)} (\tau,z).
\hspace{1cm}
\left(\any \nu, \om \in \La \right).
\label{sflow hAppell u}
\end{eqnarray}
We also note
\begin{equation}
\hf^{(k)}_{u+\gamma}(\tau, z+\gamma) = e^{-4\pi i k \frac{\gamma_2}{\tau_2}(z-u) }\,
\hf^{(k)}_u(\tau,z).
\hspace{1cm}
\left(
\any \gamma \in \frac{1}{2k} \bz \tau + \br
\right)
\label{sflow hAppell u 2}
\end{equation}
It would be difficult to show 
the last identity \eqn{sflow hAppell u 2} 
directly from the original definition \eqn{hAppell}.

~


Several remarks are in order;

\begin{description}

\item[1.]
Let us assume the rational level $k= N/K$, $N,K \in \bz_{>0}$.
In the similar manner to the analysis given in \cite{ES-NH} for the $u=0$ case, one can make the  `character decomposition' of 
regularized partition function
$\Z_{\msc{reg}}(\tau,z, \bar{z},u ; \ep)$, and extract 
the modular completions $\hchid $ \eqn{hchid u}.  Although necessary computations are slightly cumbersome,  
one can achieve the expected result;
\begin{eqnarray}
\hspace{-5mm}
\Z_{\msc{reg}}(\tau,z, \bar{z},u ; \ep) & = & e^{2\pi \frac{\hc}{\tau_2} z_1^2}\,
\sum_{\stackrel{v+K(a_L+a_R) \in N\bz}{v,a_L,a_R \in \bz_N}}\,
\hchid^{(N,K)}(v,a_L;\tau,z,u)\, \overline{\hchid^{(N,K)}(v,a_R ;\tau,z,u)}
\nn
&& 
\hspace{2cm}
+[\mbox{contribution of the continuous characters}].
\label{ch exp Zreg u}
\end{eqnarray}
Note here that (see \eqn{WI})
$$
\left. 
 \overline{\hchid^{(N,K)}(v,a_R ;\tau,z,u)} \right|_{\bar{z}=\bar{u}} = \delta^{(N)}_{a_R,0} \, e^{i\pi \frac{\hc}{\tau_2}u_2 \bar{u} }.
$$
Thus, \eqn{ch exp Zreg u} correctly reproduces the elliptic genus \eqn{EG cigar formula u} 
by setting $\bar{z}=\bar{u}$ and dividing it by the anomaly factor \eqn{anomaly factor u}.

~


\item[2.]
The `twisted elliptic genus' 
$\cZ_{\om}(\tau,z,u)$ can be also defined by the replacement \\
$
e^{-\frac{\pi k}{\tau_2} \left|\mu \right|^2 } ~ \longrightarrow ~  e^{-\frac{\pi k}{\tau_2} \left|\mu + \om  \right|^2 }
$
~ in the path-integral expression \eqn{EG cigar u}. 
We can analyze $\cZ_{\om}(\tau,z,u)$ in a parallel manner 
as the $u=0$ case and find that 
\begin{eqnarray}
\cZ_{\om}(\tau,z,u) & = & \sum_{\nu\in \La}\, e^{2\pi i \frac{1}{\tau_2} 
\Im (\om \bar{\nu}) }\, \tcZ^{(\infty)}_{\nu} (\tau,z,u)
\nn
& \equiv & \vth_u(\tau,z) \, 
\sum_{\nu\in \La}\, e^{2\pi i \frac{1}{\tau_2} 
\Im \left[ \left( \om - \frac{1}{k} u \right) \bar{\nu}\right] } 
\, \frac{\rho^{(\frac{1}{k})}(\nu-u, z)}{z-u+\nu}.
\label{twisted EG nh-E u}
\end{eqnarray}
Based on this result, we can also derive the modular transformation formulas   
\begin{eqnarray}
&& \hspace{-1.5cm}
\cZ_{\om} (\tau+1,z,u ) = \cZ_{\om} (\tau,z,u), \hspace{1cm}
\cZ_{\frac{\om}{\tau}}\left(-\frac{1}{\tau}, \frac{z}{\tau},  
\frac{u}{\tau}\right) 
= e^{i\pi \frac{\hc}{\tau} z^2}\, 
\cZ_{\om}(\tau,z,u),
\label{modular twisted EG 1 u}
\\
&& \hspace{-1.5cm}
\tcZ^{(\infty)}_{\nu} (\tau+1,z,u) = \tcZ^{(\infty)}_{\nu} (\tau,z,u), \hspace{1cm}
\tcZ^{(\infty)}_{\frac{\nu}{\tau}}\left(-\frac{1}{\tau}, \frac{z}{\tau},
\frac{u}{\tau} \right) 
= e^{i\pi \frac{\hc}{\tau} z^2}\, 
\tcZ^{(\infty)}_{\nu}(\tau,z,u).
\label{modular twisted EG 2 u}
\end{eqnarray}

~

\item[3.]
Recall that the parameter $u$ expresses the general spin structure of world-sheet fermions. 
However, one should keep it in mind that 
\begin{eqnarray}
s^{(\frac{\hc}{2})}_{-u} \cdot \cZ (\tau,z) & \neq& \cZ(\tau,z,u), 
\end{eqnarray}
for generic $u$,  despite the relation \eqn{EG infty u 2}.
This feature comes from the non-commutativity 
of the spectral flow operator $s^{(*)}_{\la}$. (See \eqn{product sflow op}
and also Appendix D).
Using the twisted elliptic genus $\cZ_{\om}(\tau,z,u)$ given above, 
we rather obtain 
\begin{equation}
s^{(\frac{\hc}{2})}_{-u} \cdot \cZ (\tau,z) = \cZ_{\frac{2u}{k}}(\tau,z,u).
\label{sflow u twisted EG formula}  
\end{equation}

~


\item[4.]
As in the $u=0$ case, we can extract the modular completions 
of extended and irreducible characters 
\eqn{hchid u}, \eqn{hchd u} from the elliptic genus \eqn{EG nh-E u}. 
We thus obtain the useful formulas
\begin{eqnarray}
\hchid^{(N,K)}(v,a;\tau,z,u) & = & \vth_u(\tau,z) \,
\sum_{b\in \bz_N}\, \sum_{\la\in a\tau+b + N\La -u } \,e^{-2\pi i \left[ \frac{b}{N}(v+Ka)- 
\frac{1}{k\tau_2} \Im (\la \bar{u})
\right]}\,
\nn
&&
\hspace{3cm}
\times
\frac{\rho^{\left(1/k\right)}(\la,z)}{z+\la},
\label{hchid nh-E u}
\\
\hchd(\nu,m;\tau,z,u)  &= &\vth_u(\tau,z) \, 
\sum_{n\in \bz}\, e^{-2\pi i \left[ \frac{n}{k}(\nu+m) - \frac{1}{k\tau_2} \Im \{ (m\tau+n)\bar{u}\}\right]}\,
\nn
&& \hspace{3cm} 
\times
\frac{\rho^{(1/k)} (m\tau+n-u , z)}{z+m\tau+n-u },
\label{hchd nh-E u}
\end{eqnarray}
which extend \eqn{hchid nh-E}, \eqn{hchd nh-E}.

\end{description}

~


\subsection{Modular Completions of Superconformal Characters}

~

As an interesting application of the inclusion of spin structure parameter $u$, let us 
discuss the modular completion of characters 
in $\mathcal N$=2 and 4 superconformal algebras.


We first look at the modular completions of the $\cN=4$ massless characters of general level $k (\in \bz_{>0})$. 

The character formula of the massless representation of isospin $\ell/2$, level $k$ ($\ell=0,1,\ldots, k$) 
is given  \cite{ET, EST} (in the $\tR$-sector) as
\begin{eqnarray}
\ch{(\stR)}{0}(k,\ell;\tau,z) 
&=&(-1)^{k-\ell} 
\frac{\th_1(\tau,z)^2}{i\eta(\tau)^3 \th_1(\tau,2z)}
\, \sum_{a=-\ell}^{\ell+1}
\, \sum_{n\in \bz}\, \frac{(yq^n)^a}{1-yq^n} q^{(k+1) n^2} y^{2(k+1) n}
\nn
&\equiv & (-1)^{k-\ell}
\frac{\th_1(\tau,z)^2}{i\eta(\tau)^3 \th_1(\tau,2z)}
\, \sum_{a=-\ell}^{\ell+1} \, 
q^{-\frac{a^2}{4(k+1)}} s^{(k+1)}_{\frac{a}{2(k+1)}\tau}\cdot f^{(k+1)}_{\frac{a}{2(k+1)}\tau}(\tau,z).
\label{N=4 ch tR} 
\end{eqnarray}
Therefore, it is natural to define its modular completion by replacing 
$f^{(k+1)}_{\frac{a}{2(k+1)}\tau}(\tau,z)$ with 
$\hf^{(k+1)}_{\frac{a}{2(k+1)}\tau}(\tau,z)$;
\begin{eqnarray}
\hch{(\stR)}{0}(k,\ell;\tau,z) 
&:=  & (-1)^{k-\ell}
\frac{\th_1(\tau,z)^2}{i\eta(\tau)^3 \th_1(\tau,2z)}
\, \sum_{a=-\ell}^{\ell+1} \, 
q^{-\frac{a^2}{4(k+1)}} s^{(k+1)}_{\frac{a}{2(k+1)}\tau}\cdot \hf^{(k+1)}_{\frac{a}{2(k+1)}\tau}(\tau,z).
\label{N=4 hch tR} 
\end{eqnarray}

However, because of the identity \eqn{sflow hAppell u 2}, 
we just obtain 
\begin{equation}
q^{-\frac{a^2}{4(k+1)}} s^{(k+1)}_{\frac{a}{2(k+1)}\tau}\cdot \hf^{(k+1)}_{\frac{a}{2(k+1)}\tau}(\tau,z)
= \hf^{(k+1)}_{0}(\tau,z).
\label{formula hf a}
\end{equation}
Hence we find 
\begin{eqnarray}
\hch{(\stR)}{0}(k,\ell;\tau,z)  
=
(-1)^{k-\ell}\frac{2(\ell+1) \th_1(\tau,z)^2}{i\eta(\tau)^3 \th_1(\tau,2z)} \hf^{(k+1)}_{0}(\tau,z)
 \equiv (-1)^{\ell}  (\ell+1) \hch{(\stR)}{0}(k,0;\tau,z) .
\label{N=4 hch tR 2} 
\end{eqnarray}
This means that after the modular completion all of the $\cN=4$ massless characters for different spins ($\ell$) collapse to a single function up to an overall constant which is just the Witten index.  
This is a somewhat striking phenomenon but we do not understand its physical significance.

Let's recall the the relation between massless characters and a massive character at 
the threshold $h=k/4$ \cite{ET,EST};
\begin{eqnarray}
&& 
\ch{(\stR)}{0}(k,\ell+1;\tau,z) + 2  \ch{(\stR)}{0}(k,\ell;\tau,z) + \ch{(\stR)}{0}(k,\ell-1;\tau,z)
 =   \ch{(\stR)}{}(k,\ell+1, h= \frac{k}{4} ;\tau,z)
\nn
&& \hspace{3cm} 
\equiv   - q^{-\frac{(\ell+1)^2}{4(k+1)}} \, \chi^{SU(2)_{k-1}}_{\ell}(\tau, 2z) \frac{\th_1(\tau,z)^2}{\eta(\tau)^3},
\hspace{1cm}
(\ell = 0, 1,  \ldots, k-1).
\label{N=4 ch rel}
\end{eqnarray}
Here, $\chi^{SU(2)_{k-1}}_{\ell}(\tau, 2z)$ denotes the spin $\ell/2$ character of $SU(2)_{k-1}$;
$$
\chi^{SU(2)_{k-1}}_{\ell}(\tau, 2z) \equiv \frac{\Th{\ell+1}{k+1}(\tau,2z) - \Th{-\ell-1}{k+1}(\tau,2z) }{i\th_1(\tau,2z)},
\hspace{1cm}
(\ell=0,1,\ldots, k-1).
$$
\eqn{N=4 hch tR 2} just  implies  
\begin{eqnarray}
&&\hch{(\stR)}{0}(k,\ell+1;\tau,z) + 2  \hch{(\stR)}{0}(k,\ell;\tau,z) + \hch{(\stR)}{0}(k,\ell-1;\tau,z)
 = 0.
\label{N=4 hch rel}
\end{eqnarray}
Thus one may formally consider that the massive characters at unitarity threshold ($h={k\over 4}$) vanish after modular completion;
\begin{equation}
\hch{(\stR)}{}(k,\ell+1, h= \frac{k}{4} ;\tau,z) \equiv 0.
\end{equation}

~


We can similarly consider the modular completion of the spectrally flowed `graviton characters' in the 
$\cN=2$ theory of $\hc \equiv 1 + \frac{2}{k}$, which is written explicitly as   
\begin{eqnarray}
\chg (n;\tau,z) &:=&   - q^{-\frac{1}{4k}} \frac{(1-q) y q^{n-1}}{(1-yq^n)(1-yq^{n-1})}\,
y^{\frac{2}{k}\left(n-\frac{1}{2}\right)}  q^{\frac{1}{k} \left(n-\frac{1}{2}\right)^2}\,
\frac{\th_1(\tau,z)}{i \eta(\tau)^3}
%
\nn
&\equiv & \chd (-1, n;\tau,z) - \chd(1,n-1;\tau,z) .
\label{ch g}
\end{eqnarray}
This corresponds to the spectral flow of identity representation with the flow momentum $n \in \bz$ in the NS-sector.
Recall the definition of the discrete character \eqn{chd u} (for the $u=0$ case);
$$
\chd(\la, n;\tau,z) \equiv \frac{\th_1(\tau,z)}{i\eta(\tau)^3} \, \frac{(yq^n)^{\frac{\la}{k}}}{1-yq^n}\, y^{\frac{2n}{k}} 
q^{\frac{n^2}{k}},
$$
where we allow $\la$ to be a general real number.

To make the modular completion of  \eqn{ch g}, one may just replace the second term $\chd(1,n-1) $ 
with $\hchd(1,n-1)$. 
However, the first term $\chd (-1, n;\tau,z)$ seems subtle, since it 
does not appear in the expansion of elliptic genus of $SL(2)/U(1)$-supercoset. 
Hence, we shall incorporate the dependence on the $u$-variable
as is the case  of the $\cN=4$ characters presented above. 
Let us note the identity 
\begin{eqnarray}
\chd (-1, n;\tau,z) & = & s^{(\frac{\hc}{2})}_{-\frac{k}{2} \tau} \cdot \chd(k-1, n ;\tau,z, u= -\frac{k}{2}\tau),
\label{formula chd -1}
\end{eqnarray}
which is immediately checked from the definition \eqn{chd u}.
Now, we already know how to modular complete the R.H.S of \eqn{formula chd -1}.
Thus it is natural to define the modular completion of the graviton character  \eqn{ch g} as
\begin{eqnarray}
\hchg (n;\tau,z) &: =& 
s^{(\frac{\hc}{2})}_{-\frac{k}{2} \tau} \cdot \hchd(k-1, n ;\tau,z, -\frac{k}{2}\tau) - \hchd(1,n-1;\tau,z) .
\label{hch g 0}
\end{eqnarray}
However, we find the identity 
\begin{equation}
s^{(\frac{\hc}{2})}_{\frac{k}{2} j \tau} \cdot \hchd(\la, n ;\tau,z, \frac{k}{2}j \tau)
= \hchd(\la, n ;\tau,z,0) \equiv \hchd(\la, n ;\tau,z) , 
\hspace{1cm} (\any j \in \bz),
\label{formula hchd j}
\end{equation}
which is similar to  \eqn{formula hf a}. 
In fact, the formula \eqn{hchd nh-E u} can be rewritten as 
\begin{eqnarray}
\hchd(\nu,m;\tau,z,u)  &= &\vth_u(\tau,z) \, 
\sum_{n\in \bz}\, e^{-2\pi i \frac{n}{k}(\nu+m) }\,
 s^{(\frac{1}{k})}_{m\tau+n} \cdot s_{-u}^{(\frac{1}{k})} \cdot
\left[\frac{e^{- \frac{\pi z^2}{k\tau_2}}}{z}\right],
\end{eqnarray}
and the commutativity of $s^{(\frac{1}{k})}_{\frac{k}{2}j\tau}$ with  
$ s^{(\frac{1}{k})}_{m\tau+n}$ readily leads to \eqn{formula hchd j}.
(See \eqn{product sflow op}.)

In this way, we have achieved the simple formula of modular completion;
\begin{eqnarray}
\hchg (n;\tau,z) & =& 
\hchd(k-1, n ;\tau,z) - \hchd(1,n-1;\tau,z) .
\label{hch g}
\end{eqnarray}
Because of the periodicity $\hchd(\la, n) = \hchd(\la+ k, n)$, one may also rewrite it 
in the form parallel to \eqn{ch g}; 
\begin{eqnarray}
\hchg (n;\tau,z) & =& 
\hchd(-1, n ;\tau,z) - \hchd(1,n-1;\tau,z) .
\label{hch g 2}
\end{eqnarray}


Similarly, the extended graviton character \cite{ES-L,ES-BH} is written as
$(a \in \bz_N)$;
\begin{eqnarray}
\chig^{(N,K)}(a;\tau,z) &:=  & 
\sum_{n\in a+ N\bz}\,\chg\left( n ; \tau, z\right)
\nn
&\equiv &
- q^{-\frac{K}{4N}} 
\sum_{n\in \bz}\,
\frac{(1-q) y q^{Nn+a-1}}{(1-yq^{Nn+a})(1-yq^{Nn+a-1})}\,
y^{\frac{2K}{N} \left(Nn+a-\frac{1}{2}\right)}  
q^{\frac{K}{N} \left(Nn+a -\frac{1}{2}\right)^2}\,
\frac{\th_1(\tau,z)}{i \eta(\tau)^3}.
\nn
&\equiv &
\chid^{(N,K)}(-K, a ;\tau,z) - \chid^{(N,K)} (K, a-1 ;\tau,z),
\label{chig}
\end{eqnarray}
and its modular completion should be 
\begin{eqnarray}
\hchig^{(N,K)}(a;\tau,z) & : = &  \sum_{n\in a+ N\bz}\,\hchg\left( n ; \tau, z\right)
\nn
&\equiv &
\hchid^{(N,K)}(N -K, a ;\tau,z) - \hchid^{(N,K)} (K, a-1 ;\tau,z)
\nn
&\equiv &
\hchid^{(N,K)}( -K, a ;\tau,z) - \hchid^{(N,K)} (K, a-1 ;\tau,z).
\label{hchig}
\end{eqnarray}
In the last line we again made use of the periodicity 
$\hchid^{(N,K)}( v, a) = \hchid^{(N,K)}( v+N , a)$.

It would be worthwhile to 
note that the `character identities'  in the $\cN=2$ case 
such as \eqn{N=4 ch rel} are written 
\begin{eqnarray}
&& \chg (n;\tau,z) - 
\chd(k-1, n ;\tau,z) + \chd(1,n-1;\tau,z)  
\nn
&& \hspace{4cm}
=  
 \frac{\th_1(\tau,z)}{i\eta(\tau)^3} \,  y^{\frac{1}{k}(2n-1)} 
q^{\frac{1}{k}n(n-1)},
\\
&& \chig^{(N,K)} (a;\tau,z) - 
\chid^{(N,K)}(N-K, a ;\tau,z) + \chid^{(N,K)}(K,a-1;\tau,z) 
\nn
&&
\hspace{4cm}
 =  q^{-\frac{K}{4N}}
 \frac{\th_1(\tau,z)}{i\eta(\tau)^3} \, 
\Th{K(2a-1)}{NK}\left(\tau, \frac{2z}{N}\right).
\end{eqnarray}
The massive characters appearing in the R.H.S again vanish after taking the modular completions;
\begin{eqnarray}
&& \hchg (n;\tau,z) - 
\hchd(k-1, n ;\tau,z) + \hchd(1,n-1;\tau,z)  
=  0,
\\
&& \hchig^{(N,K)} (a;\tau,z) - 
\hchid^{(N,K)}(N-K, a ;\tau,z) + \hchid^{(N,K)}(K,a-1;\tau,z) =0.
\end{eqnarray}

\vskip2cm

{\bf Achnowledgement}
T.E. would like to thank K.Bringmann for a discussion.
Research of T.E. is supported in part by JSPS KAKENHI grant no. 25400273,
22224001 and 23340115.

Y.S. would like to thank S.~K.~Ashok S.~Murthy, B.~Pioline, and J.~Troost for valuable discussions. 
He also thanks the organizers of International Workshop on `Mock Modular Forms and Physics' at IMSc, India, for April 14-18, 2014
for kind hospitality,  where part of this work was done. 
Research of Y.S. is supported in part by JSPS KAKENHI grant no. 23540322.


\newpage

\section*{Appendix A: ~ Notations and Useful Formulas}

\setcounter{equation}{0}
\def\theequation{A.\arabic{equation}}


~

In Appendix A we summarize the notations adopted in this paper and related useful formulas.
We assume $\tau\equiv \tau_1+i\tau_2$, $\tau_2>0$ and 
 set $q:= e^{2\pi i \tau}$, $y:=e^{2\pi i z}$;


~

\begin{description}

\item[\underline{Theta functions :}]
%
%
 \begin{equation}
 \begin{array}{l}
 \dsp \th_1(\tau,z)=i\sum_{n=-\infty}^{\infty}(-1)^n q^{(n-1/2)^2/2} y^{n-1/2}
  \equiv 2 \sin(\pi z)q^{1/8}\prod_{m=1}^{\infty}
    (1-q^m)(1-yq^m)(1-y^{-1}q^m), \\
 \dsp \th_2(\tau,z)=\sum_{n=-\infty}^{\infty} q^{(n-1/2)^2/2} y^{n-1/2}
  \equiv 2 \cos(\pi z)q^{1/8}\prod_{m=1}^{\infty}
    (1-q^m)(1+yq^m)(1+y^{-1}q^m), \\
 \dsp \th_3(\tau,z)=\sum_{n=-\infty}^{\infty} q^{n^2/2} y^{n}
  \equiv \prod_{m=1}^{\infty}
    (1-q^m)(1+yq^{m-1/2})(1+y^{-1}q^{m-1/2}), \\
 \dsp \th_4(\tau,z)=\sum_{n=-\infty}^{\infty}(-1)^n q^{n^2/2} y^{n}
  \equiv \prod_{m=1}^{\infty}
    (1-q^m)(1-yq^{m-1/2})(1-y^{-1}q^{m-1/2}) .
 \end{array}
\label{th}
 \end{equation}
\begin{eqnarray}
 \Th{m}{k}(\tau,z)&=&\sum_{n=-\infty}^{\infty}
 q^{k(n+\frac{m}{2k})^2}y^{k(n+\frac{m}{2k})} .
 \end{eqnarray}
 We use abbreviations; $\th_i (\tau) \equiv \th_i(\tau, 0)$
 ($\th_1(\tau)\equiv 0$), 
$\Th{m}{k}(\tau) \equiv \Th{m}{k}(\tau,0)$.
 We also set
 \begin{equation}
 \eta(\tau)=q^{1/24}\prod_{n=1}^{\infty}(1-q^n).
 \end{equation}
%
%
The spectral flow properties of theta functions are summarized 
as follows;
\begin{eqnarray}
 && \th_1(\tau, z+m\tau+n) = (-1)^{m+n} 
q^{-\frac{m^2}{2}} y^{-m} \th_1(\tau,z) , \nn
&& \th_2(\tau, z+m\tau+n) = (-1)^{n} 
q^{-\frac{m^2}{2}} y^{-m} \th_2(\tau,z) , \nn
&& \th_3(\tau, z+m\tau+n) = 
q^{-\frac{m^2}{2}} y^{-m} \th_3(\tau,z) , \nn
&& \th_4(\tau, z+m\tau+n) = (-1)^{m} 
q^{-\frac{m^2}{2}} y^{-m} \th_4(\tau,z) , 
\nn
&& \Th{a}{k}(\tau, 2(z+m\tau+n)) = 
e^{2\pi i n a} 
q^{-k m^2} y^{-2 k m} \Th{a+2km}{k}(\tau,2z)~.
\label{sflow theta}
\end{eqnarray}

~


We use the next symbol
\begin{equation}
\vth(\tau,z) := \frac{\th_1(\tau,z)}{\del_z \th_1 (\tau,0)} 
\equiv \frac{\th_1(\tau,z)}{2\pi \eta(\tau)^3}.
\label{def vth}
\end{equation}
Note that 
$$
\lim_{z\rightarrow 0} \left[ \del_z \vth(\tau,z) \right] =1.
$$
The modular and spectral flow properties of $\vth(\tau,z)$ are summarized as 
\begin{eqnarray}
\vth(\tau+1,z) &=& \vth(\tau, z),
\label{T vth}
\\
\vth\left(-\frac{1}{\tau}, \frac{z}{\tau}\right) &=& \frac{e^{i\pi \frac{z^2}{\tau}}}{\tau} \vth(\tau,z),
\label{S vth}
\\
\vth(\tau, z+a\tau+b) &=& (-1)^{a+b} q^{-\frac{1}{2} a^2} y^{-a} \vth(\tau,z).
\label{sflow vth}
\end{eqnarray} 
Namely, $\vth(\tau,z)$ is a weak Jacobi form of weight $-1$ and index $\frac{1}{2}$.

~

The next identity is quite useful for our calculations;
\begin{eqnarray}
&&\frac{\th_1(\tau,u+z)}{\th_1(\tau,u)} = \frac{\th_1(\tau,z)}{i\eta(\tau)^3}\,
\sum_{n\in \bz}\, \frac{w^n}{1-yq^n}, 
\nn
&& 
\hspace{2cm} \left(y\equiv e^{2\pi i z}, ~ w \equiv e^{2\pi i u}, ~ \tau_2 > 0,  ~ 
0 <  \frac{u_2}{\tau_2} < 1 \right).
\label{theta id}
\end{eqnarray}
See {\em e.g.} \cite{ES-NH} for the proof.


~

\item[\underline{Spectral flow operator :}]
(see also \cite{Eich-Zagier})
\begin{eqnarray}
s^{(\kappa)}_{\la} \cdot f(\tau,z) &:= & e^{2\pi i \frac{\kappa}{\tau_2} \la_2 \left(\la + 2z\right) }\,
f(\tau, z+\la)
\nn
& \equiv &q^{\kappa \al^2} y^{2\kappa \al} e^{2\pi i \kappa \al \beta}\,
f(\tau,z+\al \tau+ \beta) ,
\nn
&& 
\hspace{3cm} 
(\la \equiv \al \tau +\beta, ~ \any \al, \beta \in \br).
\label{def sflow op}
\end{eqnarray}

An important property of the spectral flow operator $s^{(\kappa)}_{\la}$ is the modular covariance, which precisely means the following:
\\
Assume that $f(\tau,z)$ is an arbitrary function with the modular property;
$$
f(\tau+1,z) = f(\tau,z), \hspace{1cm} 
f\left(-\frac{1}{\tau}, \frac{z}{\tau}\right) = e^{2\pi i \frac{\kappa}{\tau} z^2} \tau^{\al} \, f(\tau,z),
$$
then, we obtain for $\any \la \in \bc$
$$
s^{(\kappa)}_{\la} \cdot f(\tau+1,z) = 
s^{(\kappa)}_{\la} \cdot f(\tau,z), 
\hspace{1cm} 
s^{(\kappa)}_{\frac{\la}{\tau}} \cdot f\left(-\frac{1}{\tau}, \frac{z}{\tau}\right) 
= e^{2\pi i \frac{\kappa}{\tau} z^2} \tau^{\al}\, s^{(\kappa)}_{\la} \cdot f(\tau,z).
$$ 

~

The next `product formula' is also  useful;
\begin{equation}
s^{(\kappa)}_{\la} \cdot s^{(\kappa)}_{\la'} = 
e^{- 2\pi i \frac{\kappa}{\tau_2} \Im (\la \bar{\la'} )} s^{(\kappa)}_{\la+\la'}
= e^{- 4 \pi i \frac{\kappa}{\tau_2} \Im (\la \bar{\la'} )} s^{(\kappa)}_{\la'} \cdot s^{(\kappa)}_{\la},
\label{product sflow op}
\end{equation}
in other words, 
$$
s^{(\kappa)}_{\al\tau+\beta} \cdot s^{(\kappa)}_{\al'\tau+\beta'} = 
e^{-2\pi i \kappa(\al \beta' - \al' \beta)} \, s^{(\kappa)}_{(\al+\al')\tau+(\beta+\beta')}
=  e^{-4\pi i \kappa(\al \beta' - \al' \beta)} \, s^{(\kappa)}_{\al'\tau+\beta'} \cdot s^{(\kappa)}_{\al \tau + \beta}.
$$
We should note that the spectral flow operators do not commute with each other in general.

~


The next  notation has been repeatedly used in the main text; 
\begin{equation}
\vth_u(\tau,z)
:= s^{(1/2)}_{-u} \cdot \vth(\tau,z) 
\equiv 
q^{\frac{1}{2}\al^2} y^{-\al} e^{i\pi \al \beta}
\vth\left(\tau,z-u \right),
\label{def vth u}
\end{equation}
which satisfies
\begin{eqnarray}
\vth_u(\tau+1,z)
&=&  
\vth_u
(\tau,z),
\nn
\vth_{u/\tau}
\left(- \frac{1}{\tau}, \frac{z}{\tau}\right)
&=&  {e^{i\pi \frac{z^2}{\tau}}\over \tau} 
\vth_u
(\tau,z),
\label{modular vth u}
\\
\vth_u
(\tau,z+m\tau+n) 
&=& (-1)^{m+n} e^{2\pi i \left(m\beta - n\al \right)}
q^{- \frac{m^2}{2}} y^{-m} \, 
\vth_u
(\tau,z).
\label{sflow vth u}
\end{eqnarray}

~


\item[\underline{Error function} :]
\begin{equation}
\erf(x) := \frac{2}{\sqrt{\pi}} \int_0^{x} e^{-t^2} dt,  \hspace{1cm} (x\in \br)
\label{Erf}
\end{equation}

The next identity is elementary but  useful;
\begin{eqnarray}
&& \sgn(\nu +0) -  \erf(\nu ) 
= \frac{1}{i\pi} 
\int_{\br - i0}\, dp \, \frac{e^{-(p^2 + \nu^2)}}{p-i\nu}.
\hspace{1cm} (\nu \in \br),
\label{id erf}
\end{eqnarray}


~

\item[\underline{weak Jacobi forms} :]

~

The weak Jacobi form \cite{Eich-Zagier} for 
the full modular group $\Gamma(1) \equiv SL(2,\bz)$
with weight $k (\in \bz_{\geq 0})$ and index
$r (\in \frac{1}{2} \bz_{\geq 0})$ 
is defined by the conditions 
\begin{description}
 \item[(i) modularity :] 
\begin{eqnarray}
 && \hspace{-1cm}
\Phi\left(
\frac{a\tau+b}{c\tau+d}, \frac{z}{c\tau+d}
\right) 
= e^{2\pi i r \frac{c z^2}{c\tau+d}} (c\tau+d)^k \, \Phi(\tau,z), ~~~
\any \left(
\begin{array}{cc}
 a& b\\
 c& d
\end{array}
\right) \in \Gamma(1).
\label{modularity J}
\end{eqnarray}
\item[(ii) double quasi-periodicity :] 
\begin{eqnarray}
 \Phi(\tau,z+m\tau+n) = (-1)^{2r(m+n)} q^{-r m^2} y^{-2r m}\, \Phi(\tau,z). 
\label{sflow J}
\end{eqnarray}
%
\end{description}
In this paper, we shall use this terminology in a broader sense. 
We allow a half integral index $r$, and more crucially,  allow non-holomorphic dependence on $\tau$,
while we keep the holomorphicity with respect to $z$ \footnote
{According to  the original terminology of \cite{Eich-Zagier},
the `weak Jacobi form' of weight $k$ and index $r$ ($k, r \in \bz_{\geq 0}$) means that $\Phi(\tau,z)$ should 
be Fourier expanded as 
$$
\Phi(\tau,z) = \sum_{n\in \bz_{\geq 0}}\, \sum_{\ell \in \bz}\, c(n,\ell) q^n y^{\ell},
$$
in addition to the conditions \eqn{modularity J} and \eqn{sflow J}.
It is called the `Jacobi form' if it further satisfies  the condition: 
$c(n,\ell) =0$ for $\any n, \ell$ s.t.  $4n r -\ell^2 <0$.
}.

\end{description}

~

~


\section*{Appendix B:~ Proof of \eqn{Z sflow formula u}}

\setcounter{equation}{0}
\def\theequation{B.\arabic{equation}}


~

In this appendix we prove the  formula \eqn{Z sflow formula u}. 
Even though this identity (for the $u=0$ case) has been already shown in \cite{orb-ncpart}
by using the  modular completion of irreducible characters,
we shall here directly derive it from the path-integral expression of elliptic genus
with an arbitrary parameter of continuous spin structure $u$.  
Of course, we can readily obtain the identity \eqn{Z sflow formula} by setting $u=0$.

We shall assume $k=N/K$ with some positive integers $N$ and $K$ {\em (not necessarily 
coprime)} for the time being. We will later take the large $N$-limit with keeping $k$ a fixed value.  
We denote $\La \equiv \bz \tau + \bz$ as in the main text.

Let's  start with the orbifolding associated with the $\bz_N$-symmetry 
acting on the angle coordinate $\theta$ of cigar as 
$
\theta \, \rightarrow \, \theta + 2\pi \frac{a}{N}, 
$
$(\any a \in \bz_N)$.
The corresponding elliptic genus is given as 
\begin{equation}
\cZ^{(N)}(\tau,z,u) = \frac{1}{N} \sum_{\gamma \in \La/N\La}\,
\cZ^{(N)}_{\gamma}(\tau,z,u),
\end{equation}
where  each twisted sector is explicitly written as 
\begin{eqnarray}
\cZ^{(N)}_{\gamma}(\tau,z,u) 
& = & e^{-\frac{\pi}{k\tau_2} \left(u^2 -|u|^2\right)} e^{i\pi \frac{u_2}{\tau_2} \left(u-2z\right) }\, \lim_{\ep\rightarrow +0}\,
k \, e^{\frac{\pi}{k \tau_2} (z-u)^2} \, 
\int_{\bc} \frac{d^2 \mu}{\tau_2}\, 
\sigma(\tau, \mu,z, \bar{u}; \ep)\,
\nn
&& \hspace{1cm}
\times
\frac{\th_1 \left(\tau, \mu +  \frac{k+2}{k}z -u \right)}
{\th_1\left(\tau, \mu + \frac{2}{k}z \right)} 
\, e^{2 \pi i (z-u) \frac{\mu_2}{\tau_2}}\,
e^{-\frac{\pi k}{\tau_2}\left| \mu+ \frac{\gamma}{N} \right|^2}.
\label{twisted EG gamma}
\end{eqnarray}
We further introduce their `Fourier transformation' by the relation as 
\begin{equation}
\tcZ^{(N)}_{\delta}(\tau,z,u) := \frac{1}{N} \sum_{\gamma \in \La/N\La} \, e^{ 2\pi i \frac{1}{N\tau_2} \Im (\delta \bar{\gamma})}\,
\cZ^{(N)}_{\gamma}(\tau,z,u), 
\hspace{1cm} (\any \delta \in \La/N\La)
\label{EG Fourier rel}
\end{equation}

Now, the next identity is obvious by definition 
\begin{equation}
\cZ(\tau,z,u) \left( \equiv \cZ_{\gamma=0}^{(N)}(\tau,z,u) \right)
= \frac{1}{N}\sum_{\delta \in \La/N\La}\,
\tcZ^{(N)}_{\delta} (\tau,z ,u).
\label{Fourier relation 1}
\end{equation}

Moreover, by using the above definitions as well as that of spectral flow operator 
 \eqn{def sflow op}, 
one can directly show the next  identity
\begin{equation}
s^{(\frac{\hc}{2})}_{\la} \cdot \cZ^{(N)}(\tau,z,u) \left(\equiv 
s^{(\frac{\hc}{2})}_{\la} \cdot \tcZ^{(N)}_{\delta=0}(\tau,z,u) \right) 
= (-1)^{n_1+n_2 + n_1 n_2} e^{\frac{2\pi i}{\tau_2}\Im(\la \bar{u})}\,  
 \tcZ^{(N)}_{\la}(\tau,z,u),
\label{Fourier relation 2}
\end{equation}
for $\any \la \equiv n_1 \tau + n_2 \in \La / N\La  $.

The identity \eqn{Fourier relation 1} holds for an arbitrarily large $N \in \bz_{>0}$ as long as 
$k$ is fixed,
since $\cZ(\tau,z,u)$ depends only on $k$. 
Therefore, taking the large $N$-limit, and   
combining the identities \eqn{Fourier relation 1} with \eqn{Fourier relation 2}, 
we eventually obtain 
\begin{eqnarray}
\cZ(\tau,z, u) &=& \lim_{\stackrel{N\,\rightarrow \, \infty}{k\equiv N/K\,:\, 
\msc{fix}}} \, 
\frac{1}{N} \sum_{\la \equiv n_1\tau+n_2 \in \La/N\La}\, (-1)^{n_1+n_2 + n_1 n_2}  e^{ - \frac{2\pi i}{\tau_2}\Im(\la \bar{u})} \, 
s^{(\frac{\hc}{2})}_{\la} \cdot  \cZ^{(N)} (\tau,z,u)
\nn
& \equiv & \sum_{\la \equiv n_1\tau+n_2 \in \La}\, (-1)^{n_1+n_2 + n_1 n_2}  e^{ - \frac{2\pi i}{\tau_2}\Im(\la \bar{u})} \, 
s^{(\frac{\hc}{2})}_{\la} \cdot  \cZ^{(\infty)}(\tau,z,u),
\label{Z sflow formula u Appendix}
\end{eqnarray}
where we used 
\begin{eqnarray}
\cZ^{(\infty)} (\tau,z,u) & :=& \lim_{\stackrel{N\,\rightarrow \, \infty}{k\equiv N/K\,:\, 
\msc{fix}}} \, 
\frac{1}{N} \cZ^{(N)}(\tau,z,u)
\nn
& =& 
e^{-\frac{\pi}{k\tau_2} \left(u^2 -|u|^2\right)} e^{i\pi \frac{u_2}{\tau_2} \left(u-2z\right) }\,
\lim_{\ep \rightarrow +0}\, 
k e^{\frac{\pi X^2}{k\tau_2}} \, \int_{\Sigma} \frac{d^2 \om}{\tau_2}\,
\int_{\bc} \frac{d^2 \mu}{\tau_2} \,  
\sigma(\tau, \mu ,X+u,\bar{u}; \ep)\,
\nn 
&& 
\hspace{2cm}
\times 
\frac{\th_1\left(\tau, \mu + \frac{k+2}{k}z \right)}
{\th_1\left(\tau, \mu  + \frac{2}{k} z \right)} 
\, e^{2\pi i X \frac{\mu_2}{\tau_2}}\, 
e^{-\frac{\pi k}{\tau_2}\left| \mu+ \om \right|^2},
\label{def EG infty u}
\end{eqnarray}
in the second line.

In this way, we obtain the desired formula \eqn{Z sflow formula u}. 
Although the above proof was under the assumption $k=N/K$,
\eqn{Z sflow formula u} is correct for $\any k \in \br$, since we can 
choose arbitrarily large integers $N$ and $K$. 
(Recall that $N$ and $K$ are not assumed to be coprime.)
~~
{\bf (Q.E.D)}

~


We finally remark the `Fourier relation' that generalizes 
\eqn{Z sflow formula u Appendix}; 
\begin{eqnarray}
\cZ_{\om}(\tau,z, u) &: =& \lim_{\stackrel{N\,\rightarrow \, \infty}{k\equiv N/K, \,
\om \equiv \frac{\gamma}{N} \,:\, 
\msc{fix}}} \,  \cZ^{(N)}_{\gamma} (\tau,z,u)
\nn
& =  & \sum_{\la \in  \La}\,  
 e^{  \frac{2\pi i}{\tau_2}\Im(\om \bar{\la})} \, 
\tcZ^{(\infty)}_{\la}(\tau,z,u),
\\
\tcZ^{(\infty)}_{\la}(\tau,z,u) & := & 
\lim_{\stackrel{N\,\rightarrow \, \infty}{k\equiv N/K, \,:\, 
\msc{fix}}} \,
\frac{1}{N} \tcZ^{(N)}_{\la}(\tau,z,u)
\nn
& = & 
 (-1)^{n_1+n_2 + n_1 n_2}  e^{ - 2\pi i \frac{1}{\tau_2}\Im(\la \bar{u})}
s^{(\frac{\hc}{2})}_{\la} \cdot \cZ^{(\infty)}(\tau,z,u) ,
\nn
&& 
\hspace{4cm}
(\la \equiv n_1\tau+n_2 \in \La).
\label{Fourier rel u Appendix}
\end{eqnarray}

~

~


\section*{Appendix C:~ Proof of \eqn{hAppell nh-E u}}

\setcounter{equation}{0}
\def\theequation{C.\arabic{equation}}

~

In this appendix we present a proof of the identity 
\eqn{hAppell nh-E u}.

We shall first rewrite the original form of $\hf^{(k)}_u(\tau,z)$ \eqn{hAppell}
by an integral formula  using the identity \eqn{id erf}
$(u\equiv \al \tau+\beta, ~ \al,\beta \in \br)$;
\begin{eqnarray}
\hf_u^{(k)}(\tau,z)
&=& 
f_u^{(k)}(\tau,z)
- \frac{1}{2 \pi i} \sum_{m,r \in \bz}\,
\int_{\br+i(2k\al-0)} dp\, 
\frac{e^{-\pi \tau_2 \frac{p^2 + (r+2k\al)^2}{k}}}{p-i(r+2k\al)}\,
\left(yw^{-1} q^m\right)^r \, y^{2k m} q^{k m^2}
\nn
&=& 
\frac{i}{2\pi}\,
\sum_{m, r\in\bz}\, \left\{ \int_{\br + i(1-0)} dp
- \int_{\br-i0} dp \, \left(yw^{-1} q^m \right)
\right\}
\,
\frac{ e^{- \pi \tau_2 \frac{p^2+(r+2k\al)^2}{k}}}{p-i(r+2k\al)}\, 
\nn
&& \hspace{5cm} 
\times \frac{\left(yw^{-1} q^m\right)^{r}}
{1-y w^{-1} q^{m}} \,
y^{2k m} q^{k m^2}.
\label{hAppell int formula}
\end{eqnarray}
Using the identity $\dsp \sum_{n\in \bz}\, e^{2\pi i n \nu} = \sum_{r \in \bz}\, \delta(\nu-r)$,
we can further rewrite \eqn{hAppell int formula} as
\begin{eqnarray}
\hf_u^{(k)}(\tau,z) &=& 
\frac{i}{2\pi}\, \sum_{m \in \bz}\, \int_{-\infty}^{\infty} d\nu \,
 \sum_{n \in\bz}\, \left\{ \int_{\br + i} dp
- \int_{\br} dp \, \left(yw^{-1} q^m \right)
\right\}
\,
\frac{ e^{- \pi \tau_2 \frac{p^2+(\nu +2k\al)^2}{k}}}{p-i(\nu+2k\al)}\, 
\nn
&& \hspace{4cm} 
\times 
e^{2\pi i n \nu}
\frac{\left(yw^{-1} q^m\right)^{\nu}}
{1-y w^{-1} q^{m}} \,
y^{2k m} q^{k m^2}.
\label{hAppell int formula 2}
\end{eqnarray} 
Here, it is not obvious whether  
the infinite $n$-summation  commutes with the $\nu$-integral as well as the $m$-summation. 
We thus define 
\begin{eqnarray}
F(\tau,z,u;\cN) & := & \frac{i}{2\pi}\, \sum_{m \in \bz}\, \sum^{\cN}_{n = -\cN}\, \int_{-\infty}^{\infty} d\nu \,
 \left\{ \int_{\br + i} dp
- \int_{\br} dp \, \left(yw^{-1} q^m \right)
\right\}
\,
\frac{ e^{- \pi \tau_2 \frac{p^2+(\nu +2k\al)^2}{k}}}{p-i(\nu+2k\al)}\, 
\nn
&& \hspace{4cm} 
\times 
e^{2\pi i n \nu}
\frac{\left(yw^{-1} q^m\right)^{\nu}}
{1-y w^{-1} q^{m}} \,
y^{2k m} q^{k m^2}
\nn
& \equiv & 
\sum_{m \in \bz}\, \sum_{n= -\cN}^{\cN}\, s^{(k)}_{m\tau+n} \cdot g^{(k)}(\tau,z,u),
\label{def F cN}
\end{eqnarray}
where we set
\begin{eqnarray}
\hspace{-2cm} 
g^{(k)}(\tau,z,u) &: =& 
\frac{i}{2\pi}\, \int_{-\infty}^{\infty} d\nu \,
\left\{ \int_{\br + i} dp
- \int_{\br} dp \, \left(yw^{-1} \right)
\right\}
\,
\frac{ e^{- \pi \tau_2 \frac{p^2+(\nu +2k\al)^2}{k}}}{p-i(\nu+2k\al)}\, 
\frac{\left(yw^{-1}  \right)^{\nu}}{1-y w^{-1} } .
\label{def g k}
\end{eqnarray}
The double integral can be explicitly carried out
in the same way as \eqn{EG infty nh}, which yields 
\begin{eqnarray}
\hspace{-5mm}
g^{(k)}(\tau,z,u) &=& \int_{-2k \al}^{1-2k \al} d\nu\,
\frac{\left(yw^{-1}  \right)^{\nu}}{1-y w^{-1} }
+ \frac{i}{2\pi} \int_{-\infty}^{\infty} d\nu \int_{-\infty}^{\infty} dp\, 
 \frac{ e^{- \pi \tau_2 \frac{p^2+(\nu +2k\al)^2}{k}}}{p-i(\nu+2k\al)} 
 \left(yw^{-1}  \right)^{\nu} \nn
&=&
\frac{i}{2\pi} \frac{\left(yw^{-1}\right)^{-2k\al}}{z-u} e^{-\frac{\pi k}{\tau_2} (z-u)^2}
\equiv \frac{i}{2\pi} \frac{ e^{-\frac{\pi k}{\tau_2} (z^2-u^2)
+ \frac{2\pi k}{\tau_2} \bar{u}(z-u)}}{z-u}.
\label{eval g k}
\end{eqnarray}
Thus, we obtain 
\begin{eqnarray}
s^{(k)}_{\la} \cdot g^{(k)}(\tau,z,u) & \equiv & 
\frac{i}{2\pi} \,
e^{2\pi i k \frac{\la_2}{\tau_2} (\la +2 z)}\,
\frac{e^{-\frac{\pi k}{\tau_2} \left\{(z+\la)^2 -u^2\right\} 
+\frac{2\pi k}{\tau_2} \bar{u} (z+\la-u)} }
{z+\la-u}
\nn
&= &
 \frac{i}{2\pi} \,
e^{\frac{\pi k}{\tau_2} \left(u^2 -|u|^2 \right)} 
 \frac{e^{-\frac{\pi k}{\tau_2} 
\left[z^2 + 2(\bar{\la}- \bar{u}) z + \left|\la -u\right|^2 - (\la \bar{u} -\bar{\la}u  ) \right]}}{z+\la-u}
\nn
& \equiv & 
 \frac{i}{2\pi} \,
e^{\frac{\pi k}{\tau_2} \left(u^2 -|u|^2 \right)} \, 
 \, \frac{\rho^{(k)}(\la-u, z) \,
e^{ 2 \pi i \frac{ k}{\tau_2} \Im (\la \bar{u}) }}{z-u +\la}.
\label{eval g k 2}
\end{eqnarray}
for $\any \la \equiv m\tau+n \in \La$.

Thanks to the Gaussian behavior $\sim e^{- \frac{\pi k}{\tau_2} \left| \la \right|^2}$
for large $|\la |$ in \eqn{eval g k 2},
we find the absolute convergence;
$$
\sum_{\la \in \La} \, \left| s^{(k)}_{\la} \cdot g^{(k)}(\tau, z,u) \right| < \infty.
$$
Therefore, we can  safely conclude that 
$$
\hf^{(k)}_u(\tau,z) = \lim_{\cN\rightarrow \infty} F(\tau,z,u ; \cN) = 
\lim_{\cN\rightarrow \infty} \, \sum_{m\in \bz} \, \sum_{n=-\cN}^{\cN}  \,  s^{(k)}_{m\tau+n} \cdot g^{(k)}(\tau, z,u)  = 
\sum_{\la \in \La} \,  s^{(k)}_{\la} \cdot g^{(k)}(\tau, z,u),
$$
which proves the desired identity \eqn{hAppell nh-E u}. ~~
{\bf (Q.E.D)}


~

~


\section*{Appendix D:~ Summary of Modular Completions with General Spin Structures}

\setcounter{equation}{0}
\def\theequation{D.\arabic{equation}}

~


In Appendix D, we summarize the definitions of  modular completions 
of the irreducible and extended characters of $\cN=2$ superconformal algebra with 
general spin structures parameterized by a continuous parameter $u \in \bc$. 
They are natural extensions of those 
given in \cite{ES-NH,orb-ncpart} for the $u=0$ case. 
We assume $\tau \in \bh$ and 
set 
$q \equiv e^{2\pi i \tau}$, $ y \equiv e^{2\pi i z}$, $  w \equiv e^{2\pi i u} \equiv e^{2\pi i (\al \tau + \beta)}$.

~


\noindent
\underline{\bf Modular Completions of Irreducible Characters : }
\begin{eqnarray}
\hchd(\la,n;\tau,z,u) &:=& e^{- \frac{\pi}{k\tau_2} \left(u^2 -\left|u\right|^2\right)} \, \vth_u(\tau,z) 
\,\sum_{\nu \in \la +2\al + k\bz}\, \left\{ \int_{\br + i(k-0)} dp\, -
\int_{\br-i0} dp \, \left(y w^{-1 }q^{n} \right) 
\right\}\,
\nn
&& \hspace{2cm} 
\times
\frac{ e^{- \pi \tau_2 \frac{p^2+\nu^2}{k}} }{p-i\nu}
\, \frac{\left(yw^{-1}q^n \right)^{\frac{\nu-2\al }{k}}}{1-yw^{-1}q^n}\, y^{\frac{2n}{k}} q^{\frac{n^2}{k}} 
\nn
& \equiv & \chd([\la]_{\al} ,n;\tau,z,u)
- 2\pi i e^{- \frac{\pi}{k\tau_2} \left(u^2 -\left|u\right|^2\right)} \, 
\vth_u(\tau,z) \,
\nn
&& \hspace{2cm}
\times
\sum_{\nu \in \la +2\al + k\bz}\,
\int_{\br-i0} dp \, \frac{ e^{- \pi \tau_2 \frac{p^2+\nu^2}{k}} }{p-i\nu}
\, \left(yw^{-1}q^n \right)^{\frac{\nu-2\al}{k}} y^{\frac{2n}{k}} q^{\frac{n^2}{k}} 
\nn
& \equiv &
(-1)^n e^{-2\pi i \beta n} \, 
s^{(\frac{\hc}{2})}_{n\tau}\cdot
s^{(\frac{\hc}{2})}_{-u} \cdot \hchd (\la+2\al,0;\tau,z),
\nn
&& 
\left(
\la \in \br, ~ n\in \bz, ~
[\la]_{\al} \equiv \la ~ (\mod k\bz), ~ -2\al \leq [\la]_{\al} \leq k-2\al \right),
\label{hchd u}
\end{eqnarray}
and the irreducible character is defined by\footnote
   {The factor $e^{- \frac{\pi}{k\tau_2} \left(u^2 -\left|u\right|^2\right)}$ might look peculiar, since 
$\chd (\la,n;\tau,z,u)$ should be holomorphic with respect to the modulus $\tau$.
However, this is indeed a natural definition from the physical viewpoint, 
as is implied in the second line of \eqn{chd u}.
Note that $e^{- \frac{\pi}{k\tau_2} \left(u^2 -\left|u\right|^2\right)} \equiv e^{-\frac{2\pi i}{k} \al (\al\tau+\beta) }$,
and the `holomorphicity' here truly means that 
$$
\left. \frac{\del}{\del \bar{\tau}} \chd (\la,n;\tau,z,\al\tau+\beta) \right|_{z,\al,\beta\,:\, \msc{fixed}} =0.
$$
 } 
\begin{eqnarray}
\chd (\la,n;\tau,z,u) &:=& 
- 2\pi i e^{- \frac{\pi}{k\tau_2} \left(u^2 -\left|u\right|^2\right)} \, 
\vth_u(\tau,z) \, 
\frac{(yw^{-1}q^n)^{\frac{\la}{k}}}{1-yw^{-1} q^n}\,
 y^{\frac{2n}{k}}  q^{\frac{n^2}{k}}\,
\nn
& \equiv & (-1)^n e^{-2\pi i \beta n} \, 
s^{\left(\frac{\hc}{2}\right)}_{n\tau}\cdot
s^{\left(\frac{\hc}{2}\right)}_{-u} \cdot \chd (\la+2\al,0;\tau,z),
\nn
&& 
\left(-2\al \leq \la \leq k-2\al, ~ n\in \bz\right).
\label{chd u}
\end{eqnarray}
Here 
$\chd (\la, n;\tau,z) \equiv \chd (\la, n ;\tau,z,u=0)$
($\hchd (\la, n;\tau,z) \equiv \chd (\la, n ;\tau,z,u=0)$)
denotes the (modular completion) of the  character associated to the $n$-th spectral flow of 
discrete irrep. generated by the Ramond vacua; 
\begin{equation}
h= \frac{\hc}{8}, ~~~ 
Q = \frac{\la}{k} - \frac{1}{2}, ~~(0\leq \la \leq k) 
\end{equation}

Note that the modular completion $\hchd(\la,n)$ has the periodicity under $\la\,\rightarrow \, \la + k$, which is obvious 
from the definition \eqn{chd u}, while $\chd(\la,n)$ does not.

~



\noindent
\underline{\bf Modular Completions of Extended Characters : }

~

We assume  $k = N/K$, ($N,K \in \bz_{>0}$), or equivalently, $ \hc = 1+ \frac{2K}{N}$. 
\begin{eqnarray}
\hspace{-1cm}
\hchid^{(N,K)}(v,a;\tau,z,u) & := & \sum_{n\in a+ N\bz}\, \hchd\left(\frac{v}{K},  n;\tau,z,u\right)
\nn
& \equiv & 
e^{- \frac{\pi}{k\tau_2} \left(u^2 -\left|u\right|^2\right)} \, \vth_u(\tau,z) 
\,\sum_{r \in v + N\bz}\, \sum_{n \in a + N\bz} \left\{ \int_{\br + i(N-0)} dp\, -
\int_{\br-i0} dp \, \left(y w^{-1}q^{n} \right) 
\right\}\,
\nn
&& \hspace{1cm} 
\times
\frac{ e^{- \pi \tau_2 \frac{p^2+(r+2K \al)^2}{NK}} }{p-i(r+2K\al)}
\, \frac{\left(yw^{-1}q^n \right)^{\frac{r}{N}}}{1-yw^{-1}q^n}\, y^{\frac{2n}{k}} q^{\frac{n^2}{k}} 
\nn
& \equiv & \chid^{(N, K)}([v]_{\al} ,a ;\tau,z,u)
- 2\pi i e^{- \frac{\pi}{k\tau_2} \left(u^2 -\left|u\right|^2\right)} \, 
\vth_u(\tau,z) 
\nn
&& 
\hspace{1cm} 
\times 
\sum_{r \in v + N\bz}\, \sum_{n \in a + N\bz} 
\, \int_{\br-i0} dp \, \frac{ e^{- \pi \tau_2 \frac{p^2+(r+2K \al)^2}{NK}} }{p-i(r+2K\al)}
\, \left(yw^{-1}q^n \right)^{\frac{r}{N}} y^{\frac{2n}{k}} q^{\frac{n^2}{k}} ,
\nn
&& 
\left(v,a\in \bz_N, ~ [v]_{\al} \equiv v ~ (\mod N), ~ -2K\al \leq [v]_{\al} < N-2K\al \right),
\label{hchid u}
\end{eqnarray}
and 
\begin{eqnarray}
\hspace{-1cm}
\chid^{(N,K)}(v,a;\tau,z,u) & := & \sum_{n\in a + N\bz}\, \chd\left(\frac{v}{K}, n ;\tau,z,u\right)
\nn
& \equiv & -2\pi i  
 e^{- \frac{\pi}{k\tau_2} \left(u^2 -\left|u\right|^2\right)} \, 
\vth_u(\tau,z) \, \sum_{n\in a+N\bz}\, 
\frac{(yw^{-1}q^n)^{\frac{v}{N}}}{1-yw^{-1} q^n}\,
 y^{\frac{2n}{k}}  q^{\frac{n^2}{k}},
\nn
&& 
\hspace{1cm} 
\left(a \in \bz_N, ~ v\in \bz, ~ -2K\al \leq v \leq N-2K\al \right).
\label{chid u}
\end{eqnarray}
We note that  $\chid^{(N,K)}(v,a;\tau,z) \equiv 
\chid^{(N,K)}(v,a;\tau,z,0)$ is the extended discrete character 
introduced in \cite{ES-L,ES-BH}. 
Again the modular completion $\hchid^{(N,K)}(v,a)$ is periodic under $v\, \rightarrow \, v+N$, while 
$\chid^{(N,K)}(v,a)$ is not.

~

We should remark that,
even though 
\begin{eqnarray}
\hchd(\la, 0;\tau,z,u) &=& s^{(\frac{\hc}{2})}_{-u} \cdot \hchd(\la+2\al,0;\tau,z), 
\nn
\chd(\la, 0;\tau,z,u) &=& s^{(\frac{\hc}{2})}_{-u} \cdot \chd(\la+2\al,0;\tau,z), 
\end{eqnarray}
holds by definition, 
we do not have  similar simple relations for $\hchd(\la, n)$ $(\chd(\la, n))$ with $n\neq 0$ and
$\hchid^{(N,K)}(v,a)$ ($\chid^{(N,K)}(v,a)$). 
This fact is due to the non-commutativity of the spectral flow operators $s^{(\kappa)}_{\la}$, 
expressed in  \eqn{product sflow op}.
In cases of $K \in 2\bz_{>0}$, and for $u\in \frac{1}{2}\bz \tau + \frac{1}{2}\bz $,
we find 
\begin{eqnarray*}
\hchid^{(N,K)}(v,a;\tau,z, u) &=& s^{(\frac{\hc}{2})}_{-u} \cdot \hchid^{(N,K)}(v,a;\tau,z),
\\
\chid^{(N,K)}(v,a;\tau,z, u) &=& s^{(\frac{\hc}{2})}_{-u} \cdot \chid^{(N,K)}(v,a;\tau,z).
\end{eqnarray*}
Namely, one can define `$\hchid^{(N,K)\, (\sigma)}(v,a)$' (`$\chid^{(N,K)\, (\sigma)}(v,a)$')
with $\sigma = \NS, \tNS,  \R$ by the 1/2-spectral flows in the standard manner 
in those cases. This aspect is consistent with the definitions of modular completions of 
$\NS$, $\tNS$, $\R$ sectors given in  \cite{ncpart-NENS5}.

~


The modular and spectral flow properties of $\hchd$, $\hchid$
are almost the same as those given in \cite{ES-NH,orb-ncpart};
\begin{eqnarray}
&& 
\hchd \left(\la,n ; \tau+1, z, u \right)
= e^{2\pi i \frac{n}{k} \left(\la+ n \right)}\,
\hchd \left(\la,n ; \tau, z , u\right),
\label{T hchd u}
\\
&& 
\hchd \left(\la,n ; - \frac{1}{\tau}, \frac{z}{\tau}, \frac{u}{\tau}\right)
= e^{i\pi \frac{\hc}{\tau}z^2}\,
 \frac{1}{k} \int_0^k d\la'  \,\sum_{n' \in \bz}\,
e^{2\pi i \frac{\la \la' - (\la+2n)(\la'+2n')}{2k}}
\, \hchd (\la',n';\tau,z,u).
\nn
&&
\label{S hchd u}
\\
&& \hchd (\la,n;\tau,z+r\tau+s,u) = (-1)^{r+s} e^{2\pi i (\beta r -\al s)} e^{2\pi i \frac{\la+2n}{k}s}
q^{-\frac{\hc}{2}r^2} y^{-\hc r}\, \hchd(\la,n+r;\tau,z,u),
\nn
&& \hspace{12cm} (\any r,s \in \bz).
\label{sflow hchd u}
\end{eqnarray}
\begin{eqnarray}
&& 
\hchid \left(v,a ; \tau+1, z, u \right)
= e^{2\pi i \frac{a}{N} \left(v+ K a \right)}\,
\hchid \left(v,a ; \tau, z , u\right).
\label{T hchid u}
\\
&& 
\hchid \left(v,a ; - \frac{1}{\tau}, \frac{z}{\tau}, \frac{u}{\tau}\right)
= e^{i\pi \frac{\hc}{\tau}z^2}\,\frac{1}{N} \, \sum_{v'=0}^{N-1} \,\sum_{a'\in \bz_N}\,
 e^{2\pi i \frac{vv' - (v+2Ka)(v'+2Ka')}{2NK}}
\, \hchid (v',a';\tau,z, u),
\label{S hchid u}
\\
&& 
\hchid (v,a;\tau,z+r\tau+s,u) = (-1)^{r+s} e^{2\pi i (\beta r -\al s)} e^{2\pi i \frac{v+2Ka}{N}s}
q^{-\frac{\hc}{2}r^2} y^{-\hc r}\, \hchid(v,a+r;\tau,z,u),
\nn
&& 
\hspace{12cm} (\any r,s \in \bz).
\label{sflow hchid}
\end{eqnarray}

~

We also note the formula for Witten indices;
\begin{eqnarray}
\hspace{-1cm}
&& \lim_{z\,\rightarrow\, u}\, \chd (\la,n;\tau,z,u) = 
\lim_{z\,\rightarrow\, u}\, \hchd (\la,n;\tau,z,u) = \delta_{n,0} \, e^{-i\pi \hc \al u },
\nn
\hspace{-1cm}
&& \lim_{z\,\rightarrow\, u}\, \chid^{(N,K)} (v,a;\tau,z,u) = 
\lim_{z\,\rightarrow\, u}\, \hchid^{(N,K)} (v,a;\tau,z,u ) =  
\delta_{a,0}^{(N)} e^{-i\pi \hc \al u }
\nn
&& \hspace{4cm}
\equiv 
\left\{
\begin{array}{ll}
e^{-i\pi \hc \al u } & ~~~ a\equiv 0~ (\mod N) \\
0 & ~~~ \mbox{otherwise}.
\end{array}
\right.
\label{WI}
\end{eqnarray}


~


\end{document}